\documentclass[11pt]{article}
\usepackage{amsmath,amsthm,latexsym,amssymb,amsfonts,epsfig}

\newtheorem{lemm}{Lemma}[section]
\newtheorem{prop}[lemm]{Proposition}
\newtheorem{theo}[lemm]{Theorem}
\newtheorem{coro}[lemm]{Corollary}
\newtheorem{defi}[lemm]{Definition}

\newcommand{\R}{\mathbb{R}}                  
\newcommand{\C}{\mathbb{C}}                  
\newcommand{\Man}{\Sigma}

\newcommand{\scpr}[2]{\left\langle\, #1\,,\, #2\, \right\rangle} 
\newcommand{\betr}[1]{\left\lvert #1 \right\rvert}     
\newcommand{\Ad}{\text{Ad}}
\newcommand{\mc}[1]{\mathcal{#1}}

\DeclareMathOperator{\supp}{supp}

\newcommand{\skripta}{\mathfrak{A}}           
\newcommand{\con}{\mathcal{A}}                
\newcommand{\gcon}{\overline{\mathcal{A}}}    
\DeclareMathOperator{\cyl}{Cyl}               

\newcommand{\hilb}{\mathcal{H}}               

\newcommand{\hPsi}{\hat{\Psi}}
\newcommand{\hX}{\hat{X}}
\newcommand{\U}{{\cal U}}

\begin{document}
\title{Uniqueness of diffeomorphism invariant
states on holonomy -- flux algebras}
\author{Jerzy Lewandowski$^{1,2}$\thanks{lewand@fuw.edu.pl}, Andrzej
Oko{\l}\'ow$^{2,5}$\thanks{oko@fuw.edu.pl}, Hanno
Sahlmann$^{1}$\thanks{hanno@gravity.psu.edu},\\ Thomas
Thiemann$^{3,4}$\thanks{tthiemann@perimeterinstitute.ca}}
\date{\small 1. Center for Gravitational Physics and Geometry,
Physics Department, 104 Davey, Penn State, University Park, PA 16802, USA\\
2. Instytut Fizyki Teoretycznej,
Uniwersytet Warszawski, ul. Ho\.{z}a 69, 00-681 Warszawa, Poland\\
3. Albert Einstein Institut, MPI f.\ Gravitationsphysik, Am
M{\"u}hlenberg 1, 14476 Golm, Germany\\
4. Perimeter Institute for Theoretical Physics and University of
Waterloo, 31 Caroline Street North, Waterloo, Ontario N2L 2Y5, Canada\\
5. Department of Physics and Astronomy, Louisiana State University, Baton Rouge, LA 70803-4001, USA\\[.5cm]
{\small  AEI-2005-093, CGPG-04/5-3}} \maketitle
\begin{abstract}
Loop quantum gravity is an approach to quantum gravity that
starts from the Hamiltonian formulation in terms of a connection
and its canonical conjugate. Quantization proceeds in the spirit
of Dirac: First one defines an algebra of basic kinematical
observables and represents it through operators on a suitable
Hilbert space. In a second step, one implements the constraints. The
main result of the paper concerns the representation theory
of the kinematical algebra: We show that there is only one
cyclic representation invariant under spatial diffeomorphisms.

While this result is particularly important for loop quantum gravity,
we are rather
general: The precise definition of the abstract $*$-algebra of
the basic kinematical observables we give could be used for any
theory in which the configuration variable is a connection with a
compact structure group. The variables are constructed from the
holonomy map and from the fluxes of the momentum conjugate to the
connection. The uniqueness result is relevant for any such theory
invariant under spatial diffeomorphisms or being a part of a
diffeomorphism invariant theory.
\end{abstract}
\section{Introduction}
In the Hamiltonian analysis of theories of gauge potentials, the
configuration space usually is the space $\con$  of connections
defined on a principal fiber bundle $\Pi:P\rightarrow \Sigma$ of a
compact structure group $G$. The cotangent bundle $T^*\con$
(appropriately defined) with the natural symplectic structure
becomes the phase space.

In addition to the Hamiltonian equations of motion, the theory will
exhibit constraint equations. The constraints play a double role
in a Hamiltonian theory. On the one hand they generate a group of
symmetries of
the phase space referred to as the gauge transformations, on the other
hand the set of solutions of the constraints defines the
constraint surface of the phase space.

The simplest example for such a kind of theory is certainly
Maxwell theory, where the structure group is $U(1)$. A more
general example is Yang-Mills theory, where the structure
group may be an arbitrary compact Lie group $G$. In this case the
group of the gauge transformations is the group of the fiber
preserving automorphisms of the given  bundle, homotopic to
the identity. The group is often referred to as the ``Yang Mills gauge
transformations''.

Another example, in fact the one which has triggered the present
investigations, is gravity, formulated in terms of real Ashtekar
variables \cite{aa1,crrev,ttrev,alrev,crbook}. In the $3+1$ case, the
structure group is $SU(2)$, the bundle is trivial and defined over
a $3$-manifold. The group of the gauge transformations generated
by the constraints contains {\it all} the bundle automorphisms
homotopic to the identity map. In terms of a local section, the
group becomes the semi-direct product of the Yang-Mills gauge
transformations and the diffeomorphisms of $\Sigma$ homotopic to
the identity map. This Hamiltonian formulation is the starting
point of the loop quantum gravity (LQG, for brevity) program.

To quantize such a theory \`a la Dirac, one first seeks
appropriate basic variables. These are preferred functions
separating the points of the phase space which are then quantized.
This part of the procedure is called \textit{kinematical}
hereafter. The constraints are then imposed as operator equations
on the kinematical Hilbert space or in an appropriately selected
dual.

In the present paper we are concerned with two issues arising in
the kinematical quantization framework of LQG and of every theory
of connections whose phase space is $T^*\con$. The first issue
concerns the choice of  basic classical variables and a definition
of a corresponding (abstract) quantum $*$-algebra. We slightly
generalize and improve details of the ideas developed in LQG  and
define a quantum $*$-algebra $\skripta$ of basic quantum
variables. Our definitions are valid for arbitrary dimension $D\geq 2$ of
the base manifold $\Sigma$, arbitrary compact structure group $G$,
and arbitrary bundle $P$.

The second issue arises when we look for representations: If
$\skripta$ admits more than one representation, which one are we
going to choose to carry out the Dirac quantization program?
Our result here will hold in a more specific setting than our
definition of the algebra $\skripta$: We will show that in the
case of {\it diffeomorphism invariant theories}\footnote{In the
non-trivial bundle case, we mean invariance with respect to a
group of automorphisms of $P$ which induces, by the bundle projection
$\Pi$, all the diffeomorphisms of $\Sigma$ homotopic with the
identity map.}, upon restricting to diffeomorphism invariant
representations, this issue will not arise: we find a
unique cyclic representation

In the following, let us explain the two results of the paper a
bit more in detail and relate them to what has already been
achieved elsewhere.

{\bfseries The classical algebra:} For the sake of informal
presentation, let us choose a (local) trivialization of the
bundle
$P$ and use the notation of field theory. (The main part of
the paper will be kept in the geometric and algebraic style.)
Then the phase space consists of pairs $(A,E)$ of fields defined
on $\Man$, where:  $(i)$ $A$ is a differential 1-form taking
values in the Lie algebra $\mathfrak{g}$ of the gauge group $G$
and $(ii)$ $E$ is a vector density of weight $1$ taking values
in $\mathfrak{g}^*$, the dual vector space to $\mathfrak{g}$. The
non-vanishing Poisson relations between the fields evaluated at
points can be written as
\begin{equation}
\label{canonical}
  \{A_a^i(x),E^b_j(y)\}=\delta_a^b\delta^i_j\delta(x,y),
\end{equation}
where  in a local coordinate system $(x^1,...,x^D)$ in $\Man$ and
in a basis $\{\tau_1,...,\tau_d\}$ of $\mathfrak{g}$ the fields
are decomposed into $A=A^i_a dx^a\otimes\tau_i$ and
$E=E^a_i\partial_a\otimes \tau^i$ ($\tau^1,...,\tau^d\in
\mathfrak{g}^*$ denote the dual basis). The first question to ask
is which functionals of $A$ and $E$ should be quantized. A very
natural answer is obtained by considering the geometric nature of
the fields $A$ and $E$: $A$ is a 1-form on $\Sigma$ and therefore
integrals of $A$ along 1-dimensional submanifolds are well
defined. $E$ on the other hand,  as a vector density of weight one
can be turned into a pseudo $(D-1)$-form $\tilde{E}$ (still
$\mathfrak{g}^*$ valued) using the totally antisymmetric symbol,
namely $\tilde{E}= \frac{1}{(D-1)!}E^a_i
\epsilon_{aa_1...a_{D-1}}dx^{a_1}\wedge...\wedge
dx^{a_{D-1}}\otimes\tau^i$. Hence it can be integrated over $(D-1)$-dimensional hyper-surfaces of $\Sigma$. Asking in addition for
simple transformation behavior of the functionals of $A$ upon a
change of trivialization, that is with respect to
$$A\mapsto g^{-1}Ag+ g^{-1}dg,\ \ E\mapsto g^{-1}Eg$$
where $g$ is an arbitrary (locally defined) $G$ valued function in
$\Man$, one is led to consider functionals depending on $A$ via
the Wilson loop functionals
\begin{equation*}
h_\alpha[A]=\mathcal{P}\left(\exp -\int_\alpha
  A\right),
\end{equation*}
where $\alpha$ is a path in $\Man$. A similar requirement
applied to the canonical conjugate field $E$ leads to the
flux-like variables
\begin{equation}E_{S,f}\ =\ \int_S\tilde{E}_if^i,\label{flux}
\end{equation}
where $S$ is a $(D-1)$-dimensional surface and $f:S\rightarrow
\mathfrak{g}$ is a function of compact support on $S$. Starting
from the bracket \eqref{canonical} these variables can be endowed
with a Lie algebra structure with a remarkable geometric flavor
which was systematically explored in \cite{Ashtekar:1997eg,ACZ}:
The functions $\Psi:\con \rightarrow \C$ depending on $A$ via the
Wilson loop functionals only form the algebra of cylindrical
functions and every flux variable $E_{S,f}$ acts as a derivation
$X_{S,f}$ on this algebra, defined by the Poisson
bracket\footnote{The Poisson bracket in (\ref{X}) preserves the
space of cylindrical functions and the Dirac delta is absorbed
completely by the integrations involved in the definitions of the
holonomy and flux. This fact was pointed out for the first time in
\cite{aa1,crflux}. The specific flux derivation used in this paper
was defined in \cite{Ashtekar:1997eg}.}
\begin{equation} X_{S,f}\Psi\ :=\ \{\Psi,E_{S,f}\}.\label{X}
\end{equation}
This is also the approach we will use in the present paper. The
product (\ref{X}) is well defined provided that the intersection
between the path $\alpha$ with the surface $S$ contains finitely
many isolated points. A simple condition that ensures this
property uses a real analytic structure on $\Man$, analytic paths
and analytic surfaces. Correspondingly, analytic diffeomorphisms
of $\Man$ are among the natural symmetries inherited from $\Man$
that define automorphisms of the algebra of the basic variables.
The analyticity  requirement, however, breaks the local character of the
non-analytic diffeomorphisms group. Therefore, the most important
difference to the treatment of the previous papers on the subject
\cite{1,2,Sahlmann:2003in,Okolow:2003pk,ol2} is that we will employ
here a considerably larger group of symmetries.
We will not require that the
diffeomorphisms we consider be analytic everywhere but, roughly speaking,
analytic only up to submanifolds of lower dimension.
Some care has to be taken in the precise definition of this notion, mainly
to insure that they form a group and that
application of these diffeomorphisms produce
surfaces and edges that still have finitely many isolated
intersection points.
The important point is that this larger symmetry group now contains local
diffeomorphisms, and this will be instrumental for proving the uniqueness
result.\footnote{A more radical enlargement of the symmetry group of
the algebra has been advocated for a long time by Zapata (see ex.
\cite{zapata}). Recently, a similar
enlargement has been implemented in \cite{rovelli}, \cite{christ2}.
See also \cite{velhinho} for a discussion of
these questions.}
A more technical difference as compared to the LQG literature is
that, following \cite{ol2},
we will be working with arbitrary space-time dimensions and not
assume a trivialization of the $G$-bundle.

{\bfseries The quantum algebra:} The next step in the quantization
program is to define the quantum algebra $\skripta$.  Stated in a
heuristic way, we want to define an abstract $*$--algebra of
quantum objects $\hat{h}_\alpha$, $\hat{X}_{S,I}$ whose relations
reflect $(i)$ the multiplicative  structure of the functions of
the Wilson loop functionals and the derivations, and $(ii)$ the
complex conjugation structure of the functions of the Wilson loop
functionals and the flux functionals. Such an algebra has been
defined in \cite{1,2,Okolow:2003pk,Sahlmann:2003in}
on various levels of rigor and abstraction.
Here, we will reach an equivalent, precise
definition by using intuition from geometric quantization.

{\bfseries Representations, uniqueness:} After one has defined the
quantum algebra $\skripta$, according to the Dirac
quantization program $\skripta$ has to be represented on a
Hilbert space, the constraints have to be implemented as
operators, and solutions to the constraints have to be found.
Generically,
$\skripta$ will admit an infinite number of inequivalent representations,
so it
is an important question which one of them is the right one to
use. Ultimately, this question can only be answered by exhibiting
one or more representations in which the program can be followed
through to the end, leading to a bona fide quantization  of the
theory.

However, there are clearly more and less natural choices of
representations to try first: Most importantly, if the classical
theory has symmetries that act on $\skripta$ by a group of
automorphisms then it is natural to try to find a representation
in which these automorphisms are unitarily implemented. A second
natural idea is to first look at irreducible or at least cyclic
representations as the simple building blocks, out of which more
complicated representations could eventually be built. Finally, if
$\skripta$ is not a Banach-algebra, one has to worry about domain
questions and it is somewhat natural to consider representations
first that have simple properties in this respect.

A simple formulation of these properties can be given by asking
for a state (i.e. a positive, normalized, linear functional) on
$\skripta$ that it is invariant under the classical symmetry
automorphisms of $\skripta$.
Given a state on $\skripta$ one can
define a representation via the GNS construction. This
representation will be cyclic by construction. Furthermore, by
construction it
has a common invariant dense domain for all the operators
representing elements of $\skripta$. Finally, if the state is
invariant under some automorphism of $\skripta$, its action is
automatically unitarily implemented in the representation.

In this article, we will investigate the class of representations
of $\skripta$ delineated above in a special case, namely if the
theory under consideration is invariant under diffeomorphisms of
the manifold $\Sigma$ in the trivial bundle case, and
automorphisms of the bundle in the general case. Most prominently,
this is the case for gravity, written in terms of connection
variables, as used in loop quantum gravity. It follows from
\cite{aldif,Ashtekar:1997eg}, that for the case of interest for
loop quantum gravity, $D=3$ and $G=\text{SU}(2)$, a state with
these properties exists. The corresponding representation has
subsequently served as a cornerstone in the LQG program. Moreover,
this representation can be immediately generalized to arbitrary
dimension and arbitrary compact gauge group. Therefore the
requirements above do not reduce considerations to the empty set.
However, it is an important question for the LQG program, and at
least an interesting mathematical question in general, whether
there exist other representations with these properties.  Our
analysis will show that this is not the case: the only state that
is invariant under the group of diffeomorphisms described above is
the one used in LQG. This is a more satisfying result than the
ones obtained in \cite{1,2,Okolow:2003pk,ol2,Sahlmann:2003in}.
However it relies heavily on the enlargement of the symmetry group
of a state not used in
earlier publications.\\
While work on this manuscript was in progress, similar results as the ones
that we will present here have been obtained in \cite{christ2}. The
technical setup of \cite{christ2} differs somewhat from the one used
here, and we refer to Section \ref{concl} for a comparison.
\section{The Holonomy-Flux $*$-algebra}
The goal of this section is a definition of the $*$-algebra
$\skripta$ of basic, quantum observables. We have already
mentioned the algebra in the introduction and explained its meaning,
in this section we will give a complete definition. As indicated,
in our approach we will base all definitions on a new category of
manifolds that is
larger than the analytic category but smaller than the $m$-times
differentiable one. The technical definitions and
proofs in this respect are relegated to the appendix. Let us
start here by giving a more intuitive description, justification
for this enlargement and outline of the properties relevant in our
paper.
\subsection{Semianalytic structures}
In this work we consider a $D$-dimensional differential manifold $\Sigma$.
The differentiability class $C^m$ is fixed,  $m\ge 1$.
Our elementary variables -- already mentioned in the introduction and
carefully defined in the following sections -- are constructed by
using curves (later called edges) and co-dimension one submanifolds
(faces) of $\Sigma$. A necessary condition
for the Poisson bracket between the variables to be finite is that every
edge intersects every face in at most finite number of isolated
intersection points plus a finite number of connected segments (i.e.
edges in themselves).
To ensure this condition we need to carefully define a class of curves and
submanifolds we consider. It will be also important that the class be
preserved by a sufficiently large subgroup of the diffeomorphisms of $\Sigma$.
`Large' means that the subgroup contains sufficiently many diffeomorphisms
that act non-trivially only within compact regions. This is not the case,
for example, for the analytic diffeomorphism group that has usually been
considered in this context.
We solve this technical issue by defining an appropriate category of
manifolds we will call semianalytic. Next, we assume that the manifold
$\Sigma$
is equipped with a semianalytic structure. Henceforth, all the local maps,
diffeomorphisms, submanifolds and functions thereon,  are assumed to be
$C^m$ and  semianalytic. A semianalytic structure is weaker than an
analytic one, therefore it can be determined on $\Sigma$ for example by
choosing an arbitrary analytic structure.

\begin{figure}
\centerline{\epsfig{file=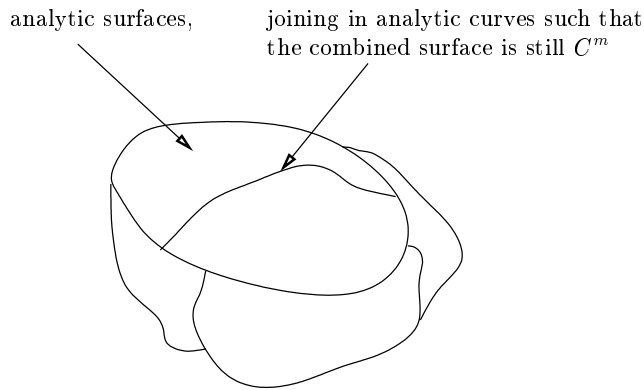}}
\caption{\label{faceexam}A semianalytic surface}
\end{figure}

Briefly, `semianalytic' means `piecewise analytic'. For example, a
semianalytic submanifold would be analytic except for on some
lower dimensional sub-manifolds, which in turn have to be piecewise analytic.
To convey the idea,
Figure \ref{faceexam} depicts a semianalytic surface in $\R^3$.
However, whereas in
the case of $\Sigma=\R$ `piecewise analytic' has a well established
interpretation, in a higher dimensional case those words admit a
huge ambiguity.
Therefore in the appendix we introduce  exact definitions and  prove relevant
properties. We heavily rely on the theory of the semianalytic sets
developed by {\L}ojasiewicz \cite{loj,bier-pier}.

The special property of the semianalytic category so relevant for us in this
paper, is that the intersection between every two connected submanifolds,
locally, is a finite union  of connected submanifolds. Of course, this is
also true in the analytic case. But the difference between the analytic and
the semianalytic structures is in the local character of the later ones.
Technically, the locality is expressed by the fact that every open covering
of $\Sigma$ admits a compatible semianalytic partition of unity.
\subsection{The cylindrical functions}
Let us recall from the appendix that by \textit{semianalytic edge} we mean
a connected, 1-dimensional semianalytic submanifold of
$\Sigma$ with 2-point boundary.
\begin{defi}
An edge is an oriented embedded 1-dimensional C$^{0}$ submanifold of
$\Sigma$ with a 2-point boundary, given by a finite
union of semianalytic edges.
\end{defi}

Over the manifold $\Sigma$ we fix a principal fiber bundle
\begin{equation}
\Pi:P\rightarrow \Sigma.
\end{equation}
The structure group of $P$  is denoted by $G$, and it is assumed
to be compact and connected. The right action of $G$ on $P$ will be
denoted in the usual way as $G\times P\ni (g,p)\mapsto R_gp\in P$.
We are assuming the bundle is semianalytic. On $P$ we consider the
space of the connections $\con$.

Given an edge $e$, a connection $A\in \con$ defines a bundle isomorphism
\begin{equation}
A(e): \Pi^{-1}(x)\rightarrow \Pi^{-1}(y)
\end{equation}
where $x$ and $y$ are the beginning and end points of $e$, and the
fibers of $P$ are considered as pullbacks of the bundle $P$. The
space $\con_e$ of all the bundle isomorphisms
$\Pi^{-1}(x)\rightarrow \Pi^{-1}(y)$ (in fact, $\con_e$ depends on
the points $x$ and $y$ only) can be mapped in a 1-1 way into $G$,
\begin{equation}\label{gauge}
\sigma :  \con_e \rightarrow  G,
\end{equation}
and the map, called a gauge map, is defined by a choice of two
points, $p_x\in \Pi^{-1}(x)$ and $p_y\in \Pi^{-1}(y)$,  and by
\begin{equation}
A_e(p_x)\ =\ R_{\sigma(A_e)}p_y.
\end{equation}
Therefore, it is determined up to the left and right
multiplication in $G$ by arbitrary elements $g,h\in G$,
corresponding to changing the points $p_x$ and $p_y$. In this way
$\con_e$ inherits every structure of $G$ which is left and right
invariant (including the topology, the differential manifold
structure, the Haar measure).

\begin{defi} A function $\Psi: \con\ \rightarrow\ \C$
is called cylindrical if there exists a finite set $\gamma=\{e_1,...,e_n\}$
of edges and a function
$\psi\in C^\infty(\con_{e_1}\times...\times\con_{e_n})$ such that for every
$A\in \con$
\begin{equation}
 \Psi(A)\ =\ \psi(A(e_1),...,A(e_n));\label{cyl}
\end{equation}
In this case, we say that $\Psi$ is compatible with $\gamma$ and $\psi$.
\end{defi}

Every cylindrical function is compatible with many sets of edges.
Without lack of generality, we may assume that $\gamma$ is an
embedded graph, that is, if two edges $e_I \not= e_J$ intersect,
then the intersection is contained in the boundary of each of them
\cite{al1}.\footnote{In \cite{al1} the analyticity was assumed.
However, owing to Proposition \ref{finiteness} the semianalyticity
assumption used in the definition of the edges is sufficient.}
The boundary points of edges constituting a graph $\gamma$ are called the
vertices of $\gamma$. It is easy to see, that all the cylindrical
functions set up a subalgebra of the algebra of all the complex
valued functions defined on $\con$; we denote it by $\cyl$. In a
natural way it admits definition of an involution and a norm
\begin{equation}
\label{csnorm}
\Psi^\star\ :=\ \bar{\Psi},\ \ \|\Psi\|\ :=\ \sup_{A\in \con}|\Psi(A)|.
\end{equation}
\subsection{The Ashtekar-Isham quantum configuration space
$\gcon$}
The space of connections is considered here a configuration space.
However, promoting the cylindrical functions to the basic position
variables on $\con$   (an over-complete set of variables) is
equivalent to embedding $\con$ into the Gel'fand spectrum of the
unital C$^*$-algebra  $\overline{\cyl}$ defined as the completion
of $(\cyl, \|\cdot\|, *)$. Elements of the Gel'fand spectrum of
$\overline{\cyl}$ have a geometric interpretation of generalized
(or distributional) connections on $\con$. We recall now
definition of the generalized connections (see \cite{velhinho} for
a recent review, and \cite{ai,al1,baez1,mm,aldif,alproj} for the
origins).

Consider the space $\cal{E}$ of all the edges  in $\Sigma$
including  the trivial one. Certain pairs $(e,e') \in {\cal
E}\times {\cal E}$ can be composed, yielding a new edge. More
precisely, let the beginning point of $e$ be the end point of
$e'$, then we define
\begin{equation}
e\circ e':= \overline{e\cup e'\setminus (e\cap e')},
\end{equation}
\textit{provided the result is again an edge}, where the line
stands for the completion, and the beginning (end) point of $e\circ
e'$ is defined to be the beginning (end) point of $e$ ($e'$). If
$e'$ differs from $e$ in orientation only, then $e\circ e'$ is
trivial, hence we will also use the notation $e^{-1}$ for the edge
$e$ with orientation reversed.

On the other hand, from the principal fiber bundle $P$, for pairs of points $x,y\in \Sigma$
one has the bundle isomorphisms $\Pi^{-1}(x)\rightarrow \Pi^{-1}(y)$,
and those from $\Pi^{-1}(x)$
to $\Pi^{-1}(y)$ can be composed with those from $\Pi^{-1}(y)$
to $\Pi^{-1}(z)$, yielding isomorphisms again.
\begin{defi}
A generalized connection ${\bar A}$ on $P$ assigns to every edge $e$ a
bundle isomorphism
\[
\bar{A}(e):\Pi^{-1}(e_s)\rightarrow\Pi^{-1}(e_t),
\]
where $e_s$ is the beginning (source) of the edge $e$, and $e_t$
is its end (target), such that
\begin{equation}
{\bar A}(e\circ e')= {\bar A}(e)\circ {\bar A}(e'),\ {\rm and}\
{\bar A}(e^{-1})=\left({\bar A}(e)\right)^{-1}
\end{equation}
whenever $e\circ e'$ is defined. We denote the space of generalized
connections by $\gcon$.
\end{defi}
Every cylindrical function $\Psi$ is naturally extendable to
$\gcon$, by using a compatible graph $\gamma$ and function $\psi$
(see (\ref{cyl})), namely
\begin{equation}
\Psi(\bar{A})\ :=\ \psi(\bar{A}(e_1),...,\bar{A}(e_n)),
\end{equation}
(the result defines a unique function on $\gcon$, independent of
choice of the compatible $\gamma$ and $\psi$.) Given any
$\bar{A}\in \gcon$, the map
\begin{equation}
\overline{\cyl}\ \ni\Psi\mapsto \Psi({\bar A})\in \C,
\end{equation}
is continuous in $(\overline{\cyl}, \|\cdot\|)$ and defines a C${^*}$-algebra
homomorphism,  that is an element of the Gel'fand spectrum.
Moreover, every element of the Gel'fand spectrum can be
represented by a generalized connection in that way
\cite{al1,aldif}. In this way we identify the spectrum with
$\gcon$.

\begin{defi}
The Ashtekar-Isham quantum configuration space for the loop
quantization of the theory of connections defined on $P$ is the
space $\gcon$ of the generalized connections.
\end{defi}

\subsection{Generalized vector fields tangent to $\gcon$}
Given a finite dimensional manifold as a configuration space, and
the cotangent bundle as the phase space,  the momenta correspond
to the tangent vector fields\footnote{Every vector field on a
manifold defines naturally a function on the cotangent bundle. If
the manifold is a configuration space, and the cotangent bundle
with the natural Poisson bracket is the phase space, then the
function is linear in `momenta'.} and there is available an
elegant geometric quantization scheme. This idea is easily
generalized to an infinite dimensional $\con$, but for the
quantization, one  would need a measure on $\con$ in our case
required to be invariant with respect to the automorphisms of
the bundle $P$.
Instead,  Ashtekar and Isham defined $\gcon$ and proposed to embed
$\con$ in $\gcon$ because the latter has naturally defined
compact topology and is therefor easier to treat. However, $\gcon$ does
not have a manifold structure. Nonetheless, the fluxes of the electric
field and the
corresponding  derivations (\ref{X}) defined in $\cyl$ do lead to
a quite precise definition of a generalized vector field tangent
to $\gcon$. We introduce it in this subsection in a geometric,
manifestly trivialization invariant way.

We  define now on $\gcon$  generalized vector fields which
correspond to the derivations (\ref{X}), that is to the smeared fluxes
of the frame field $E$. The generalized vector fields are labeled
by {\it faces}, and appropriate  {\it smearing functions}.
A face $S$ is introduced in the appendix (see Definition \ref{face})
as a co-dimension 1 submanifold of $\Sigma$, oriented in the sense
that the normal bundle of $S$ is equipped with an orientation.
 Now, we will carefully define the smearing functions.
Our emphasis is on the geometric, gauge invariant characteristics,
and on careful specification of the class of fields we are going to
use.
Let $S$ be a face.  Consider the bundle
\begin{equation}
P_S\ :=\ \Pi^{-1}(S)\subset P
\end{equation}
equipped with the principal fiber bundle structure induced by the
bundle $P$.
\begin{defi}\label{smearing}
Given a face $S$, a smearing vector field is  a compactly supported
semianalytic vector field  defined on the bundle $P_S$,
tangent to the fibers of the bundle and invariant under
the  action of the structure group $G$ in $P$.
\end{defi}
Let $f$ be a smearing vector field on $P_S$. Denote by $\exp(\cdot\,f)$ the
corresponding flow. The map
\begin{equation}
\exp(tf):P_S\rightarrow P_S
\end{equation}
assigned by the flow to each  $t\in \R$  preserves the fibers of $P_S$ (i.e.
$\Pi\circ \exp(tf) = \Pi$) and commutes with the right action of the structure group (i.e. $\exp(tf)R_g=R_g \exp(tf)$, for every $g\in G$). It is easy to show, that
the flow is semianalytic:  in a local trivialization of $P_S$, the vector
field $f$ corresponds to an element of the Lie algebra of $G$ and the flow
can be expressed by the usual exponential map.
Restricted to each fiber $\Pi^{-1}(x)\subset P_S$, the flow  becomes a
fiber automorphism $\exp(tf)_x$,
\begin{equation}
\exp(tf)_x\ :=\ \exp(tf)\vert_{\Pi^{-1}(x)}.
\end{equation}
We use the latter one to define  below a  1-dimensional
group formed by maps $\theta^{(t)}:\ \gcon\ \rightarrow \gcon$
which, briefly speaking, give every generalized connection
${\bar A}$ a `translation' $\exp(\pm tf)$ supported on those edges
which intersect
the face $S$ transversally (in the topological sense) where the
sign depends on the orientation of the edge with respect to the orientation
of $S$.  To define it, note that $S$  admits an open
neighborhood  ${\cal U}\subset \Sigma$ such that
$${\cal U}\setminus S = {\cal U}^-\cup {\cal U}^+$$
where ${\cal U}^-$ and ${\cal U}^+$ are disjoint, each of them is open in
$\Sigma$, connected and non-empty. The labels `$+$' and `$-$' correspond
to the orientation of $S$.
An action of the generalized flow $\theta^{(t)}$ on a generalized
connection $\bar{A}\in \gcon$,  can be defined by
using only a subclass of edges taken into account in what follows:
\begin{equation}\label{flow}
\theta^{(t)}({\bar A})(e)\ :=\
\begin{cases}{\bar A}(e)\exp(\frac{1}{2}tf)_x&\text{ if $e\cap S = \{x\}$
and $e\setminus x \subset {\cal U}^+$}\\
             {\bar A}(e)\exp(-\frac{1}{2}tf)_x&\text{ if $e\cap S = \{x\}$
             and $e\setminus x\subset {\cal U}^-$}\\
{\bar A}(e)&\text{if $e\cap S = \emptyset$  or  $e\cap \bar{S} =
e$}
\end{cases}
\end{equation}
where $x$ stands for the beginning point of $e$. Every edge $e'$ can
be written as a composition of edges of the type given on the right
hand side of equation (\ref{flow}) and their inverses, therefore
$\theta^{(t)}({\bar A})(e')$ is determined by \eqref{flow} and the
requirement that $\theta^{(t)}$ maps generalized connections to
generalized connections. Given an orientation of $S$, the resulting
flow is independent of the choice of the neighborhoods ${\cal
U},{\cal U}^-,{\cal U}^+$.

Importantly, the pullback $\theta^{(t)*}$ preserves $\cyl$ and for
every cylindrical function $\Psi$, the derivative
\begin{equation}
X_{S,f}\Psi\ := \frac{d}{dt}\Psi(\theta^{(t)}({\bar
A}))\,\vert_{t=0}\,\label{Xvec}
\end{equation}
is a well defined  element of $\cyl$. This definition is
equivalent to (\ref{X}) and it is its manifestly trivialization
independent version. An important observation is that the action
of the $X_{S,f}$ on the cylindrical functions is linear in the vector
field $f$, i.e. $X_{S,f_1+f_2}=X_{S,f_1}+X_{S,f_2}$.
The explicit formula for the action of the
operator can be found for example in \cite{alrev}.

\begin{defi}\label{Xdef} The operator $X_{S,f}:\cyl\rightarrow \cyl$
defined in (\ref{Xvec}), where $S$ is a face and $f$  is a
smearing vector field (see Definition \ref{smearing})
will be called the flux vector field corresponding to $(S,f)$.
\end{defi}

The space of the linear combinations of the operators
$\cyl\rightarrow \cyl$ of the form
\begin{equation}
\Psi\cdot X_{S_1,f_1}, \ \ \Psi\cdot [X_{S_1,f_1},X_{S_2,f_2}],\ \
\Psi\cdot[...[X_{S_1,f_1},X_{S_2,f_2}],...,X_{S_k,f_k}],
\end{equation}
where $\Psi\in\cyl$ and $X_{S_1,f_1}, X_{S_2,f_2},\ldots$ are the
flux vector fields will be called the space of generalized vector fields
tangent to $\gcon$, and denoted by $\Gamma(T\gcon)$.
It will be also convenient to use the complexification
$\Gamma(T\gcon)^{(\C)}$ of $\Gamma(T\gcon)$. Note, that every
$Y\in \Gamma(T\gcon)^{(\C)}$ is a derivation in $\cyl$, that is
\begin{equation}
Y(\Psi\Psi')\ =\ Y(\Psi)\Psi' + \Psi Y(\Psi').
\end{equation}

Continuing the analogy with the geometric quantization in the
finite dimensional case, consider the vector space
\begin{equation}
\skripta_{\rm class}\ :=\ \cyl\times \Gamma(T\gcon)^{(\C)},
\end{equation}
equipped with:
\begin{itemize}
\item the Lie bracket $\{\cdot, \cdot\}$,
\begin{equation}
\{(\Psi,Y),\,(\Psi',Y')\}\ :=\ -(Y(\Psi')-Y'(\Psi),\, [Y,Y']),
\end{equation}
\item the complex conjugation  $\ \bar{{}}\ $ defined by the complex
conjugations in $\cyl$ and  in $\Gamma(T\gcon)^{(\C)}$
 extended to a map
\begin{equation}
\skripta_{\rm class}\ni (\Psi,Y)=a \mapsto
\bar{a}:=(\bar{\Psi},\bar{Y})\in \skripta_{\rm class}
\end{equation}
(in $\Gamma(T\gcon)^{(\C)}$ the c.c. is defined naturally
as $\bar{Y}(\Psi):=\overline{Y(\bar{\Psi})}$).
\end{itemize}

\begin{defi} The classical Ashtekar-Corichi-Zapata
holonomy-flux algebra is the Lie algebra $(\skripta_{\rm class},\,
\{\cdot,\cdot\})$ equipped with $\ \bar{{}}\ $ as involution.
\end{defi}

The ACZ algebra $\skripta_{\rm class}$ admits also an action of
the the algebra $\cyl$
\begin{align*}
\cyl\times \skripta_{\rm class}\ &\rightarrow \skripta_{\rm class},\\
(\Psi', (\Psi,Y))\ &\mapsto \Psi'\cdot (\Psi,Y)\ :=\
(\Psi'\Psi,\Psi'Y).
\end{align*}
\subsection{The quantum $*$-algebra}
The ACZ classical holonomy-flux Lie algebra $\skripta_{\rm
class}$, is used now as a set of labels to define an abstract
$*$-algebra. Consider  the $*$-algebra of the finite formal linear
combinations of all the finite sequences of elements of
$\skripta_{\rm class}$ with the obvious vector space structure,
the associative product $\cdot$, and involutive anti-linear
algebra anti-isomorphism $*$, defined, respectively, as follows
\begin{align}
(a_1,...,a_n)\cdot (b_1,...,b_m)\ &=\ (a_1,..., a_n, b_1,...,
b_m)\\
(a_1,...,a_n)^*\ &=\ (\overline{a_n},...,\overline{a_1}).
\end{align}
Divide the algebra by a two sided ideal defined by the following
elements (consisting of 1-element and 2-element sequences):
\begin{align}
(\alpha a) - \alpha (a)&,\ (a+b)\ - (a)-
(b),\label{vec}\\
(a,b) - (b,a) &- i(\{a,b\}),\label{q}\\
(\Psi,a)&-(\Psi a)\label{com},
\end{align}
given by all $\alpha\in \C$, $a,b\in {\skripta}_{\rm class}$ and
$\Psi\in \cyl$.  The first class (\ref{vec}) of elements of the
ideal relates the linear structure of $\skripta_{\rm class}$ with
the linear structure of the resulting quotient. The second class
(\ref{q}) of elements encodes the familiar quantum relation between
the bracket $\{\cdot,\cdot\}$ in $\skripta_{\rm class}$ and the
commutators in the quantum algebra $\skripta$. The third
class\footnote{The theorem we formulate and prove in this paper uses
only the fact that the two sided ideal contains elements of the
third class for $a\in\cyl$.} (\ref{com}) encodes  the module
structure of the ACZ Lie algebra $\skripta_{\rm class}$ over the
algebra $\cyl$. Shorter, the algebra $\skripta$ may be also viewed
as the algebra $\exp(\otimes\skripta_{\rm class})$ divided by the
identities (\ref{q}) and (\ref{com}). Note that each of the classes
(\ref{vec},\ref{q},\ref{com}) is preserved by $*$. Denote the
quotient $*$-algebra by ${\skripta}$.

\begin{defi}
The  quantum holonomy-flux $*$-algebra is the (unital) $*$-algebra
$(\skripta, *)$.
\end{defi}

The classical ACZ algebra $\skripta_{\rm class}$ is naturally mapped
in $\skripta$,
\begin{equation}\label{embedding}
\skripta_{\rm class}\ \rightarrow\ \skripta
\end{equation}
in the sense that $\skripta$ is isomorphic to the enveloping algebra
of $\skripta_{\rm class}$ (see \eqref{vec}, \eqref{q}), divided by
additional identities \eqref{com} to preserve  the structure of a
$\cyl$-module. The images in ${\skripta}$ of
 1-element sequences $((\Psi,0))$ or $((0,Y))$ where $\Psi\in\cyl$ and $Y\in
\Gamma(T\gcon)^{(\C)}$ will be denoted by $\hat{\Psi}$ and $\hat{Y}$
respectively. They generate the algebra $\skripta$. In particular,
for every cylindrical function $\Psi$ and every flux vector field
$X_{S,f}$,
\begin{equation}
\hat{\Psi}^*\ =\ \hat{\bar{\Psi}},\ \ \ \hat{X}_{S,f}^*\ =\
\hat{X}_{S,f}.
\end{equation}
It is easy to see that the map (\ref{embedding}) is an embedding.
Here is a simple argument\footnote{We thank Wojtek Kami\'nski for
help.}. Consider a representation $\pi'_0:\exp(\otimes\skripta_{\rm
class})\rightarrow {\rm L}(\cyl)$, where ${\rm L}(\cyl)$ is the
algebra of the linear maps $\cyl\rightarrow\cyl$, defined on the
generators as follows,
\begin{equation}
\pi'_0((\Psi,X))\Phi \ =\ \Psi\Phi -i\{\Phi,X\}.
\end{equation}
It is easy to check, that each of the elements (\ref{q}) and
(\ref{com}) is in the kernel of $\pi'_0$. Therefore, $\pi'_0$ passes
naturally to a representation $\pi_0:\skripta \rightarrow {\rm
L}(\cyl)$,
\begin{equation}
\pi_0\ :\ \skripta\ \rightarrow {\rm L}(\cyl).
\end{equation}
The point is, that the composition of the maps (\ref{embedding})
 and $\pi_0$,
\begin{equation}
\skripta_{\rm class}\ \rightarrow\ \skripta\ \rightarrow\ {\rm
L}(\cyl),
\end{equation}
is obviously injective. Hence the first map is also injective.

\subsection{The elements of $\skripta$}
Owing to the identities in $\skripta$ defined by the third class
(\ref{com}) of elements defining the ideal above, the image
$\widehat{\cyl}\subset \skripta$ of $\cyl$ upon  the map
(\ref{embedding})
\begin{equation}
\cyl\ni\Psi\ \mapsto \hat{\Psi} \in \skripta
\end{equation}
is a $*$-subalgebra. Due to \eqref{vec}, the map is linear, it is
multiplicative due to \eqref{com}, and bijective as the restriction
of (\ref{embedding}) and hence a $*$-isomorphism  between $\cyl$ and
$\widehat{\cyl}$. Every element of the algebra $\skripta$ is a
finite linear combination of elements of the form
\begin{equation}\label{decomp}
\hPsi,\  \hPsi_1 \hX_{S_{11},f_{11}},\
\hPsi_2  \hX_{S_{21},f_{21}} \hX_{S_{22},f_{22}},\  ... \
\hPsi_k  \hX_{S_{k1},f_{k1}} ...
\hX_{S_{kk},f_{kk}},\ ...\
\end{equation}
where $\Psi,\Psi_i\in \cyl$ and $X_{S_{ij},f_{ij}}$ are the flux
vector fields for all the $i,j= 1,...,k$. For example,
\begin{equation}
a=\hX_{S,f}\hPsi \hX_{S',f'}\ =\
-i\widehat{X_{S,f}(\Psi)}\hX_{S',f'}\ +\
\hPsi\hX_{S,f}\hX_{S',f'}.
\end{equation}
\subsection{Symmetries of $\skripta$}
The group of  the semianalytic automorphisms of the principal
fiber bundle $P$ acts naturally in the space $\con$ of
connections. The action preserves the algebra $\cyl$ of the
cylindrical functions, the norm $\|\cdot\|$ and the $*$
involution. Therefore it induces an action of the  bundle
automorphism group  in the space $\gcon$ of generalized
connections. The action of the bundle automorphism group on the
flux vector fields can be viewed either as the action on operators
$X_{S,f}:\cyl\rightarrow\cyl$ (\ref{X}), or as the action on the
field $E$ and its flux functional (\ref{flux}). Both definitions
are equivalent and lead to an appropriate action of the bundle
automorphisms on the labels, i.e. the faces and the smearing
vector fields. In this way, the bundle automorphism group induces
an isomorphism of the ACZ classical Lie algebra $\skripta_{\rm
class}$, and finally a $*$-isomorphism of  the quantum $*$-algebra
$\skripta$. In this subsection we discuss the action of the
automorphisms/diffeomorphisms in detail. But before doing that let
us make a remark on the relation between the bundle $P$
automorphisms and the manifold $\Sigma$ diffeomorphisms. For every
bundle automorphism
\begin{equation}
\tilde{\varphi}:P\rightarrow P
\end{equation}
there is a unique diffeomorphism
\begin{equation}
\varphi: \Sigma \rightarrow \Sigma
\end{equation}
such that
\begin{equation}
\Pi\circ\tilde{\varphi} = \varphi\circ \Pi\label{diffs}.
\end{equation}
In our case both of them are semianalytic. If the diffeomorphism $\varphi$
is the identity map, then the corresponding automorphism is fiber preserving,
and we can call it a Yang-Mills gauge transformation. On the other hand, all
the $\Sigma$ diffeomorphisms homotopic to the identity map are related to
the bundle automorphisms via (\ref{diffs}). In this sense, the bundle
automorphisms represent also the diffeomorphisms of $\Sigma$.

Now we turn to the technical details of the action of the bundle automorphism
group in the quantum $*$-algebra $\skripta$.

For every edge (only the end
points of $e$ matter here), the map $\tilde{\varphi}$ defines the
following map \cite{ol2}
\begin{align}
\Ad_{\tilde{\varphi}}:\con_e &\rightarrow \con_{\varphi(e)}\nonumber\\
A(e)\  &\mapsto\ \tilde{\varphi}\circ
A(e)\circ\tilde{\varphi}^{-1}.
\end{align}
The map extends naturally to the product space,
\begin{align}
\Ad_{\tilde{\varphi}}:\con_{e_1}\times...\times\con_{e_N}\
&\rightarrow\
\con_{\varphi(e_1)}\times...\times \con_{\varphi(e_N)}\nonumber\\
(\,A(e_1),...,A(e_N)\,)\  &\mapsto\
(\Ad_{\tilde{\varphi}}(A(e_1)),...,
\Ad_{\tilde{\varphi}}(A(e_N))\,).
\end{align}
It will be relevant later that the map is smooth which can
easily be seen by fixing any trivialization of the fibers of $P$ in
question.  Via $\Ad$, the automorphism $\tilde{\varphi}$ acts in
the space of the generalized connections
\begin{align}
\overline{\Ad}_{\tilde{\varphi}}:&\quad\gcon\ \rightarrow \gcon\label{Ad}\\
&\left(\overline{\Ad}_{\tilde{\varphi}}{\bar A}\right)(e)\ :=\
\Ad_{\tilde{\varphi}}({\bar A}(\varphi^{-1}(e))).
\end{align}
The pullback $\overline{\Ad}_{\tilde{\varphi}}^*$ preserves the
space of the cylindrical functions. Indeed, for every cylindrical
function $\Psi$ given by (\ref{cyl})
\begin{equation}
(\overline{\Ad}_{\tilde{\varphi}}^*\Psi)({\bar A})\ =\
(\Ad_{\tilde{\varphi}}^*\psi) ({\bar A}(\varphi^{-1}(e_1)),...,
{\bar A}(\varphi^{-1}(e_n)))
\end{equation}
meaning that $\overline{\Ad}_{\tilde{\varphi}}^*\Psi$ is
compatible with the graph $\varphi^{-1}(\gamma):=
\{\varphi^{-1}(e_1),...,\varphi^{-1}(e_n)\}$ and the function
$\Ad_{\tilde{\varphi}}^*\psi\in
C^\infty(\con_{\varphi^{-1}(e_1)}\times...\times
\con_{\varphi^{-1}(e_n)})$.

Given a flux vector field $X_{S,f}$, the map
$\overline{\Ad_{\tilde{\varphi}}}$ defined above maps the
generalized flow (\ref{flow}) into a new flow
$\overline{\Ad_{\tilde{\varphi}}}{\theta}^{(\cdot)}$. It
is easy to check, that the new flow is the flow of the flux vector
field $X_{\varphi(S), \tilde{\varphi}_*f}$ \cite{ol2},
hence
\begin{equation}
{\overline{\Ad_{\tilde{\varphi}}}}_*X_{S,f}\ =\
{X_{\varphi(S), \tilde{\varphi}_*f}}
\end{equation}

Finally, the natural action of $\tilde{\varphi}$ on $\skripta$,
\begin{equation}
\alpha_{\tilde{\varphi}}
: \skripta\rightarrow \skripta,
\end{equation}
is a $*$-algebra automorphism determined by the following action
on the generators $\hat{\Psi}$ and $\hat{X}_{S,f}$, where
$\Psi\in\cyl$ and $X_{S,f}$ are arbitrary:
\begin{equation}\label{alpha}
\alpha_{\tilde{\varphi}}\hat{\Psi}\ :=\
\widehat{\overline{\Ad}_{\tilde{\varphi}^{-1}}^*\Psi}, \ \
\alpha_{\tilde{\varphi}}\hat{X}_{S,f}\ :=\ \hat{X}_{\varphi(S),
\tilde{\varphi}_* f}.
\end{equation}
The action satisfies
\begin{equation}
\alpha_{\tilde{\varphi_1}\circ \tilde{\varphi_2}}\ =\
\alpha_{\tilde{\varphi_1}}\circ \alpha_{\tilde{\varphi_2}}.
\end{equation}

 It should be pointed out, that  the quantum holonomy-flux
$*$-algebra $\skripta$ can potentially admit more symmetries. The
relevance of  the bundle automorphisms lies in the fact that they
are the symmetries of a diffeomorphism invariant classical theory.

\section{States, GNS.}
\label{statesGNS}
In all of the following, we will be concerned with states $\omega$ on
$\skripta$ and their GNS representations. Recall, that

\begin{defi} A state on a $*$-algebra $\skripta$ is a functional
$\omega:\skripta\rightarrow\C$, such that for every $\alpha\in
\C$, and every $a,b\in\skripta$
\begin{align}
\omega(\alpha a+b)\ =\ \alpha\omega(a) +\omega(b),\ \
\omega(a^*)\ =\ \overline{\omega(a)}\label{state1}\\
\omega(a^*a)\ \ge 0\ \ \ \ \ \ \ \ \omega(I)\ =\ 1\label{state2},
\end{align}
where $I$ stands for the unity element of $\skripta$.
\end{defi}
Given a state on $\skripta$, we can construct the corresponding
GNS representation ($\hilb_\omega,$ $\pi_\omega,$ $\Omega_\omega,$)
where $\hilb_\omega$ is a Hilbert space, $\pi_\omega$ a
representation of $\skripta$ on $\hilb_\omega$ and $\Omega_\omega$
a vector in $\hilb_\omega$ which, when viewed as a state on
$\skripta$, coincides with $\omega$. A detailed exposition of the
GNS construction for algebras of unbounded operators can be found
for example in \cite{schmud}. Here we will only need the following
elements and properties that are easy to prove:
$(i)$ the linear space of the equivalence classes
\begin{equation}
\left[\skripta \right]\ :=\ \skripta/\mathfrak{I}
\end{equation}
where $\mathfrak{I}$ is the left ideal formed by all $a\in\skripta$ such
that $\omega(a^*a)=0$, is equipped by the state $\omega$ with the following
product
\begin{equation*}
 \scpr{[a]}{[b]}\ :=\ \omega(a^*b),
\end{equation*}
where for every $a\in \skripta$, $[a]\in \skripta/\mathfrak{I}$
stands for the equivalence class defined by $a$; $(ii)$  the
product provides a norm $\|a\|_\omega=\sqrt{ \scpr{[a]}{[a]}}$ in
$\left[ \skripta\right]$, and the completion
\begin{equation}
\hilb_\omega :=\ \overline{\left[ \skripta \right]}
\end{equation}
together with the product $ \scpr{\cdot}{\cdot}$ is a Hilbert
space; $(iii)$ to every element $a$ of $\skripta$ we assign a
linear but in general unbounded operator $\pi_\omega(a)$ acting in
$\left[\skripta\right]$
\begin{equation}
\pi_\omega(a)[b]\ :=\ [ab],\ \  {\rm for \ \ every\ \ } \
b\in\skripta;
\end{equation}
$(iv)$ the action $\pi_\omega$  preserves the subspace
$\left[\skripta\right]$, hence $\left[\skripta\right]$
serves as a common, dense domain for all the  operators
$\pi_\omega(a),\ \ a\in\skripta$. $(v)$ The  representation
$\pi_\omega$ satisfies
\begin{equation}
 \scpr{\pi_\omega(a)[b]}{[c]}\ =\  \scpr{[b]}{\pi_\omega(a^*)[c]},
\end{equation}
for every $a,b,c\in \skripta$.

As we explained in the previous section, $\skripta$ contains the
subalgebra $\widehat{\cyl}\subset \skripta$ isomorphic as a
$*$-algebra with the algebra $\cyl$. Therefore, every state
$\omega$ defined in $\skripta$, restricted to $ \widehat{\cyl}$
defines a state on $\cyl$. On the other hand, there is known a
powerful  characterization of states defined on the completion
$\overline{\cyl}$. Fortunately, that characterization applies also
to all the states on the $*$-algebra $\cyl$ (and hence
$\widehat{\cyl}$), due to the following fact:
\begin{lemm}\footnote{This lemma is a modification of similar well
known results see for example \cite{RS}, p. 106,107.
We include it for the completeness. The factor $2$
in the inequality \eqref{omega-2} can be probably lowered to $1$, but this is not
relevant in our paper.}\label{le1}
Suppose that $\omega:\cyl\rightarrow \C$  satisfies for every
$\Psi,\Psi'\in\cyl$ and $\alpha\in\C$ the following equalities and
inequality,
\begin{gather*}
\omega(\alpha \Psi+\Psi')\ =\ \alpha\omega(\Psi) +\omega(\Psi'),\ \
\omega(\bar{\Psi})\ =\ \overline{\omega(\Psi)},\\
\omega(\bar{\Psi}\Psi)\ \ge\ 0, \ \ \ \omega(I)=1.
\end{gather*}
Then,
\begin{equation}
|\omega(\Psi)|\ \le\ 2\|\Psi\|.
\label{omega-2}
\end{equation}
Therefore, $\omega$ is continuous with respect to the norm
$\|\cdot\|$ and determines a unique extension to a state defined
on the C$^*$-algebra $\overline{\cyl}$.
\end{lemm}

\begin{proof} For every $\Psi\in\cyl$ we have
\begin{equation}
|\omega(\Psi)|\ =\ |\omega(\Psi_R)+i\omega(\Psi_I)|\ \le\
|\omega(\Psi_R)| + |\omega(\Psi_I)|,
\end{equation}
where $\Psi_R$ and $\Psi_I$ are the real and imaginary parts of
$\Psi$, respectively. Let $\Psi'\neq 0$ be $\Psi_R$ or $\Psi_I$ (if both $\Psi_R$ and $\Psi_I$ are zero then \eqref{omega-2} is satisfied trivially). For
every real number $q$ the following equality holds
\begin{equation}
\omega(\Psi')\ =\ q\|\Psi'\|- \omega( q\|\Psi'\|
I-\Psi').\label{es}
\end{equation}
Let $q>1$. Then, the function $ q\|\Psi'\|-\Psi'$ is strictly
positive, and
\begin{equation}
\con\ni A\mapsto \Psi''(A)\ :=\ \sqrt{q\|\Psi'\| -\Psi'(A)}
\end{equation}
 is a well defined function
on the space of connections. Importantly, this is also a cylindrical function.
Indeed, if we represent $\Psi'$ by a compatible graph $\gamma$ and function
$\psi'$ (see (\ref{cyl})), then the function $q\|\psi'\| -\psi'$ is
everywhere strictly positive, because the natural map $\con\rightarrow
\con_{e_1}\times...\times\con_{e_n}$ is onto. Therefore the function
\begin{equation}
\psi''\ :=\sqrt{q\|\psi'\| -\psi'}
\end{equation}
is C$^{\infty}$ and the corresponding cylindrical function is exactly $\Psi''$. But this means that the second term on the right hand side of the equality
(\ref{es}) (including the sign) is negative. Indeed,
\begin{equation}
-\omega( q\|\Psi'\|-\Psi')\ =\
-\omega(\bar{\Psi}''\Psi'')\le 0.
\end{equation}
This observation completes the proof of Lemma \ref{le1}.
\end{proof}
\section{The uniqueness theorem}
As explained in the introduction, of particular importance are
states invariant with respect to the automorphisms of the principal
fiber bundle $P$. A state $\omega$ defined on the algebra $\skripta$
is invariant with respect to a bundle automorphism
$\tilde{\varphi}:P\rightarrow P$, if for every $a\in \skripta$,
\begin{equation}
\omega(a)\ =\ \omega(\alpha_{\tilde{\varphi}}a).
\end{equation}
If $\omega$ is invariant with respect to
all the fiber preserving automorphisms, we call it Yang-Mills gauge
invariant.

\begin{defi}
If a state defined on the quantum $*$-algebra  is invariant with respect to
all the bundle automorphisms of $P$ that induce, via the bundle
projection $\Pi$, diffeomorphisms homotopic to the identity, we call
it Yang-Mills gauge
and diffeomorphism invariant or, if there is no danger of confusion,
just invariant.
\end{defi}

Given a $*$-algebra and a symmetry group,  assuming the existence of a
diffeomorphism invariant state is a strong condition.
However,  in our case one invariant state is  already known,
we recall it below. It will be, therefore, natural to ask if there
are other states with that property.

\medskip
\noindent{\bf Example:} {\it A Yang-Mills gauge invariant and diffeomorphism invariant state on $\skripta$.}
Define the action of $\omega_0$ on the elements of
$\skripta$ of the form  $a\cdot\hat{Y}$, where  $a\in \skripta$
and $Y\in \Gamma(T\gcon)^{(\C)}$ as simply
\begin{equation}\label{aX}
\omega_0(a\cdot\hat{Y})\ :=\ 0
\end{equation}
for every $a$ and every  vector field $Y$.
 To define the action of $\omega_0$ on an element $\hat{\Psi}$ corresponding
to $\Psi\in \cyl$, recall a general form (\ref{cyl}) of a cylindrical function.
Recall also, that each factor $\con_e$ in the domain
$\con_{e_1}\times ... \times\con_{e_n}$ has all the left and right invariant
structures of the bundle structure group $G$. One of them is the probability
Haar measure $\mu_{e}$. We use it to set
\begin{equation}\label{Psi}
\omega_0(\hat{\Psi})\ :=\ \int_{\con_{e_1}\times ... \times \con_{e_n}}
\psi d\mu_{e_1}\otimes ...\otimes d\mu_{e_n}.
\end{equation}
Importantly, this integral is independent of choice of the graph $\gamma$
compatible with a given $\Psi$. Due to the general form of
$a\in \skripta$ given by  (\ref{decomp})  the equalities
(\ref{aX}, \ref{Psi}) determine
a state $\omega_0$. The positivity of $\omega_0$ amounts to the positivity
of the Haar measure $\mu_{e_1}\otimes ... \otimes \mu_{e_n}$ on
$C^{\infty}(\con_{e_1}\times \ldots \times\con_{e_n})$ which is obviously true.

The state $\omega_0$ is Yang-Mills gauge and diffeomorphism invariant.
To see this, note that every automorphism
$\tilde{\varphi}$ of the bundle $P$  maps every  flux vector field
into another flux vector field, therefore the condition (\ref{aX})
is manifestly invariant. To see the invariance of the part
(\ref{Psi}) of the definition, consider  a graph $\gamma$ and
function $\psi\in C^\infty(\con_{e_1}\times ... \times\con_{e_n})$
compatible with $\Psi$  (see (\ref{cyl})), and a gauge map
${\sigma^{-1}}_\gamma:G^n\rightarrow \con_{e_1}\times ...
\times\con_{e_n}$ defined by the gauge maps (\ref{gauge}) and any
choice of points $p_x\in \Pi^{-1}(x)$ for every vertex $x$ of
$\gamma$. Then the definition (\ref{Psi}) reads
\begin{equation}
\omega_0(\Psi)\ =\ \int_{G^n}{\sigma^{-1}}_\gamma^*\psi d\mu_H,
\end{equation}
where $\mu_H$ is the probability Haar measure on $G^n$. On the
other hand, the transformed function
$\overline{\Ad}_{\tilde{\varphi}}\Psi$ is compatible with a graph
$\varphi^{-1}(\gamma)$, and the function
$\psi'=\Ad_{\tilde{\varphi}}\psi$.
 If we use for the graph  $\varphi^{-1}(\gamma)$ the gauge map
$\sigma_{\varphi^{-1}(\gamma)}$ given by the points
$\tilde{\varphi}^{-1}(p_x)$, then simply
\begin{equation}
{\sigma^{-1}}_{\varphi^{-1}(\gamma)}^*\psi'\ =\ \
{\sigma^{-1}}_\gamma^*\psi.
\end{equation}

The state $\omega_0$ is well known, and is extensively used in the loop
quantization \cite{al1,alrev}.
\medskip

Given the example of an invariant state above, let us now
state and prove our uniqueness result:

\begin{theo}
There exists exactly one Yang-Mills gauge invariant and diffeomorphism
invariant state on the quantum holonomy-flux $*$-algebra $\skripta$.
\label{main}
\end{theo}
\begin{proof}
The existence is known, see the example, therefore it suffices to prove
the uniqueness.

We will assume from now on that $\omega$ is a diffeomorphism invariant
state on $\skripta$, label the corresponding representation
obtained from the GNS construction by $(\mc{H}_\omega,\pi_\omega)$
and use the notation introduced in Section \ref{statesGNS}.
To simplify the reading, we will break down the proof
into two parts. The first of these is rather technical. It
will establish a proof of the following fundamental lemma:
\begin{lemm}
\label{fun}
Let $\omega$ be an invariant
state on $\skripta$. Then for every flux vector field $X_{S,f}$,
where $S$ is a face, and  $f$ a smearing vector field,
in the corresponding GNS-representation
\begin{equation}\label{lemm}
[\hat{X}_{S,f}]\ =\ 0.
\end{equation}
\end{lemm}
Once the lemma is established, the rest of the proof of the uniqueness is
fairly straightforward.
\begin{proof}[Proof of Lemma \ref{fun}]

Let $S$ be a face in $\Sigma$ and $f$ be a smearing vector field.
We will decompose $f$ into a certain finite sum,
\begin{equation}
f\ =\ \sum_I\sum_if_{Ii}
\end{equation}
such that each   term  $f_{Ii}$ is a smearing vector field itself
which satisfies
\begin{equation}\label{omegafiI}
[\hat{X}_{S,f_{Ii}}]\ =\ 0.
\end{equation}
Then, (\ref{lemm}) follows automatically from the linearity
\begin{equation}
X_{S,f}\ =\  \sum_I\sum_iX_{S,f_{Ii}}.
\end{equation}

To each point $x\in\Pi(\supp f)\subset \Sigma$ choose an open
neighborhood $\mc{U}_x$ in $\Sigma$  in such a way that there exists
a trivialization $\mc{T}_x$ of $\Pi^{-1}(\mc{U}_x)$,
\begin{equation}
\mc{T}_x:  \mc{U}_x\times G\ \rightarrow\  \Pi^{-1}(\mc{U}_x)\label{triv}
\end{equation}
 and such that there
is a chart $\chi_x$ containing $\mc{U}_x$ in its domain with
\begin{equation}
\chi_x(S\cap\mc{U}_x)=
\{(x^1,\ldots,x^D)\,\vert\,x^D=0,0<x^1<1,\ldots, 0<x^{D-1}<1\}.\label{chart}
\end{equation}
Since the support of $f$ is compact in $P$, we can choose from that covering a
finite subcovering $\{\mc{U}_{I}\}_{I=1}^N$ of
$\Pi(\supp f)$. We denote the corresponding trivialization by $\mc{T}_I$
and the corresponding chart by $\chi_I$.
Let $\phi_I:\Sigma\rightarrow \R$,
where $I=1,\ldots, N$, be a family of functions such that
$\supp \phi_I\subset \mc{U}_I$ for every $I$, and for every $x\in \supp f$
\begin{equation}
\sum_{I=1}^N \phi_I(x)\ =\ 1.
\end{equation}
We use that partition of unity, to decompose the smearing vector field $f$,
\begin{equation}
f\ =\  \sum_{I=1}^N  f_I, \ \ \ \ \ \ f_I\ :=\ \phi_I\,f.
\end{equation}
Each  $f_I$ ($I=1,\ldots, N$) is still a smearing vector field in the sense of
Definition (\ref{smearing}) and additionally has the appropriate support
property: $\Pi(\supp f_I)\subset\mc{U}_I$.

Now we fix $I$ and decompose the smearing vector field $f_I$ further.
Suppose $R_1$ be a vector field defined on $G$ and right invariant.
It defines naturally a vector field on  $\mc{U}_I\times G$. That vector field
is mapped  by the trivialization $\mc{T}_I$ into a vector field defined in  $\Pi^{-1}(\mc{U}_I)$,
tangent to the fibers and invariant with respect to the group action.
Let $R_i$, $i=1,....,{\rm dim}\,G$, be a basis in the vector space of the
right invariant vector fields defined on $G$. Then, every smearing
vector field $f_I$ defined on $P_S=\Pi^{-1}(S)$ is a sum of the vector fields
proportional to the vector fields $\mc{T}_{I*}R_i$, $i=1,...,{\rm dim}\,G$,
\begin{equation}\label{hi}
f_I=\sum_{i}f_{Ii},\ \ \ \ f_{Ii}= (\Pi^*h_i)\, \mc{T}_{I*}R_i,
\end{equation}
where  each coefficient $\Pi^*h_i$ is a function $h_i:S\rightarrow \R$
lifted to the bundle $P_S=\Pi^{-1}(S)$. Obviously
$\supp\,h_i\subset \mc{U}_I $. Now we can finish the proof of Lemma \ref{fun}, by
showing that necessarily (\ref{omegafiI}) is true.
To this end, fix indices $I,i$, and for every compactly supported function
$h:S\cap \mc{U}_I\rightarrow \R$
consider the following smearing vector field defined on $S$
(no summation with respect to $i$)
\begin{equation}
w(h)\ :=\  (\Pi^*h)\, \mc{T}_{I*}R_i.
\end{equation}
Consider the following product $(\,\cdot\,|\,\cdot\,)$ which given a pair of
compactly supported functions $h,g:S\cap\mc{U}_I\rightarrow \R$, assigns the
following number $(h|g)$,
\begin{equation*}
(h|g)\ :=\  \scpr{[\hX_{S,w(h)}]}{[\hX_{S,w(g)}]}.
\end{equation*}
The product $(\,\cdot\,|\,\cdot\,)$  has the following properties:

$(i)$ it is bilinear and symmetric,

$ii)$ it is invariant under diffeomorphisms
of $\Sigma$ which are supported in $\mc{U}_I$ and preserve $\mc{U}_I$
as well as $S$,

$iii)$ for $h=g=h_i$ of (\ref{hi}), it is exactly the norm squared
of the Hilbert space $\mc{H}_\omega$ element
$[\hX_{S,f_{Ii}}]$ under consideration,
\begin{equation}
\| [\hX_{S,f_{Ii}}] \|^2\ =\  \scpr{[\hX_{S,f_{Ii}}]}{[\hX_{S,f_{Ii}}]}  \ =\
(h_i|h_i).
\end{equation}

The property $ii)$ above, follows from the fact, that
via the trivialization $\mc{T}_I$, every diffeomorphism
  $\varphi:\Sigma\rightarrow \Sigma$ preserving
$\mc{U}_I$ defines  an automorphism $\tilde{\varphi}$ of the
bundle $P$ which preserves each of the vector fields
$\mc{T}_{I*}R_i$.  Therefore, if $\varphi$ additionally preserves $S$, then
the action of $\tilde{\varphi}$
(\ref{alpha}) on $\hX_{S,w(h)}$ amounts to
\begin{equation}
\alpha_{\tilde{\varphi}}\hX_{S,w(h)}\ =\
\hX_{S,w(h\circ\varphi^{-1})}.
\end{equation}

We will now show that properties i) and ii) already imply that
that $(h|h)$ is zero for every function $h:S\rightarrow \R$ with
a compact support in
$\mc{U}_I$. Then iii) shows that we have reached our goal.

Let us use the chart $\chi_I$ to push forward the action arena
into $\R^D$:
\begin{align}
\mc{U}'_I\ &:=\ \chi_I(\mc{U}_I),\ \ \ \ \ \ \ S'\ =\ \chi_I(S\cap\mc{U}_I)\\
h'\ &:=\ h\circ\chi^{-1}: S'\rightarrow \R
\end{align}
where $h'$ has a compact support and  $S'$ is defined by (\ref{chart}).

We want to extend $h'$ to a function defined in $\mc{U}'_I$
and of a compact support. Therefore, we choose an arbitrary
 semianalytic function $\kappa':\R\rightarrow\R$ such that $\kappa'(0)=1$ and
the function
\begin{equation}
(x^1,\ldots,x^D)\ \mapsto\ h'(x^1,\ldots,x^{D-1})\kappa'(x^D)
\end{equation}
has  compact support contained in ${\mc{U}'}_I$. Using these ingredients,
we can define a map\footnote{We will use here a modification of the trick
mentioned in Appendix of \cite{lm}.}
$\varphi'_\lambda:\R^D\rightarrow\R^D$, where $\lambda$ is a real
parameter, by
\begin{equation}
    {\varphi}'_\lambda(x^1,\ldots,x^D)\ :=\
(x^1+\lambda {h'}(x^1,\ldots,x^{D-1})
    \kappa'(x^D),x^2,\ldots,x^D).
\end{equation}
\begin{lemm}\label{dif}
There is $\lambda_0>0$ such that for every $0<\lambda<\lambda_0$,
$\varphi'_\lambda$ is a semianalytic diffeomorphism of $\R^D$ equal to
the identity outside of ${\mc{U}}'_I$ and preserving ${\mc{U}}'_I$.
\end{lemm}
\begin{proof}
The Jacobian of ${\varphi'}_\lambda$ is a triangular matrix and the
determinant turns out to be simply $1+\lambda \kappa'\partial_1 h'$.
Since $\lambda \kappa'\partial_1 h'$ has compact support and is
semianalytic, it is in particular bounded, and thus there is a
$\lambda_0>0$ such that $1+\lambda \kappa'\partial_1 h' >0 $ for
every $0<\lambda<\lambda_0$. Hence ${\varphi'}_\lambda$ is locally a
diffeomorphism, provided $0<\lambda<\lambda_0$. It is also a global
diffeomorphism, because outside of the support of $\kappa' h'$ it
acts as the identity and thus $\lim_{\betr{x}\rightarrow
\infty}\betr{\varphi'_\lambda(x)}=\infty$. Then a well known theorem
by Hadamard proves the assertion. Because all the functions used in
the construction of ${\varphi'}_\lambda$ are assumed to be
semianalytic, and all the operations used preserve the
semianalyticty (see Appendix), ${\varphi'}_\lambda$ is also
semianalytic. Finally, note that every bijection which is an
identity on a certain subset, necessarily preserves the complement.
\end{proof}

Now let us choose a semianalytic function $H'$ with support in
${\mc{U}'}_I$ such that
\begin{equation}
{H'}(x^1,\ldots,x^D)\ =\ x^1\ \ \ \ {\rm whenever}\ \ \ \
(x^1,\ldots,x^D)\in \supp \kappa' h'.
\end{equation}

Such a function can be easily constructed by using appropriate
partition of the unity. Let us see how each of the diffeomorphisms
${\varphi'}_\lambda$ acts on $H'$: Because of the properties of $H'$
in relation to the support of $tail\kappa' {h'}$ we find
\begin{align*}
{\varphi'}_\lambda^*{H'}(x^1,\ldots,x^D)&=
\begin{cases}
x^1+\lambda {h'}(x^1,\ldots,x^{D-1})
    {\kappa'}(x^D)\qquad &\text{on }\supp \kappa'{h'}\\
H'(x^1,\ldots,x^D) \qquad &\text{otherwise}
\end{cases}\\
&= {H'}(x^1,\ldots,x^D)+\lambda {h'}(x^1,\ldots,x^{D-1})
    {\kappa'}(x^D).
\end{align*}
Now let us pull this relation back to $\mc{U}_I$ and the manifold $\Sigma$
again by using the chart $\chi_I$. Denote the pullbacks of the functions
$H'$, $h'$, $\kappa'$ and $\varphi'_\lambda$, respectively, by
$H$, $h$, $\kappa$ and $\varphi_\lambda$.
The functions $H$ and $h\kappa$   have support
contained in  $\mc{U}_I$, therefore we can extend  them as identically
zero to the rest of $\Sigma$. Similarly, ${\varphi}_\lambda$, for every
$0<\lambda<\lambda_0$,  is a diffeomorphism  defined locally
in $\mc{U}_I$ that can be extended as the identity to the rest of
$\Sigma$, and the result is a diffeomorphism of $\Sigma$.
The above relation then reads
\begin{equation}
\varphi_\lambda^*H=H+\lambda\kappa h.
\end{equation}
Now we compute
\begin{equation}\label{tric}
  (H|H)\overset{ii)}{=}(\varphi_\lambda^*H|\varphi_\lambda^*H)=(H|H)+
\lambda(h|H)+\lambda(H|h)+\lambda^2(h|h)
\end{equation}
where we have used  the invariance of the product
$(\,\cdot\,|\,\cdot\,)$ under diffeomorphisms homotopic to the
identity (the $\varphi_\lambda$ obviously are) and the fact that
$\kappa\rvert_S=1$.
Since the equality (\ref{tric}) holds for every value of $\lambda$
provided $0<\lambda<\lambda_0$, we conclude  that
\begin{equation}
(h|h)\ =\ 0.
\end{equation}
Then, as announced above for $h=h_i$, we get the desired result, and
in turn conclude that $[\hX_{S,f}]=0$ as a vector in the GNS-Hilbert
space $\mc{H}_\omega$.
\end{proof}

Now that we have established the fundamental Lemma \ref{fun} asserting that
$[\hat{X}_{S,f}]=0$ for any face $S$ and any smearing vector field $f$ in
any GNS-representation coming from the invariant state $\omega$, we can
show that the structure of the GNS-Hilbert space $\mc{H}_\omega$ is
actually very simple.

Let us start by reminding the reader of the form
\eqref{decomp} of elements of $\skripta$ whose linear span is
$\skripta$.  It follows
immediately that dense set of vectors in $\mc{H}_\omega$
is given by the linear
span of all the vectors of the form
\begin{equation}
\begin{split}
[\hPsi],\,  \pi_\omega(\hPsi_1) [\hX_{S_{11},f_{11}}],\,
\pi_\omega(\hPsi_2  \hX_{S_{21},f_{21}}) [\hX_{S_{22},f_{22}}],
\,\ldots\\
\,\ldots,\,\pi_\omega(\hPsi_k
\hX_{S_{k1},f_{k1}}\ldots)[\hX_{S_{kk},f_{kk}}],\,\ldots
\end{split}
\end{equation}
But because of lemma \ref{fun}, of these vectors, only the ones of the
form $[\hPsi]$ are non-zero.

Therefore, all the information on the state $\omega$ is contained
in the corresponding state defined on the algebra $\widehat{\cyl}$,
\begin{equation}\label{omegacyl}
\widehat{\cyl}\ni \hat{\Psi}\mapsto \omega (\hat{\Psi})=\scpr{[\hat{I}]}{[\hat{\Psi}]}_{\hilb_\omega},
\end{equation}
where $\hat{I}$ is the unit element of $\skripta$.
Now we make use of Lemma \ref{le1} from Section \ref{statesGNS}: the
state (\ref{omegacyl}) is actually continuous with respect to the
C$^*$-norm on $\widehat{\cyl}$. Thus, using the representation
theorem by Riesz
and Markow, there exists a measure $\mu$ on $\gcon$ such that
\[
\omega(\hat{\Psi})=\int_{\gcon} \Psi\,d\mu.
\]

Notice now that Lemma \ref{fun} implies what follows 
\begin{align*}
\int_{\gcon} \overline{\Psi} X_{S,f}(\Psi')\,d\mu &=
\scpr{[\hat{\Psi}]}{[\widehat{X_{S,f}(\Psi')}]}_{\hilb_\omega}\\
&=i\scpr{[\hat{\Psi}]}{[\hat{X}_{S,f}\hat{\Psi}']-
[\hat{\Psi}'\hat{X}_{S,f}]}_{\hilb_\omega}\\
&=i\scpr{[\hat{\Psi}]}{[\hat{X}_{S,f}\hat{\Psi}']}\\
&=i\scpr{[\hat{X}_{S,f}\hat{\Psi}]}{[\hat{\Psi}']}\\
&=- \int_{\gcon} \overline{X_{S,f}(\Psi)} \Psi'\,d\mu.
\end{align*}
Setting $\Psi=I$ (i.e. the constant function on $\gcon$ of the value $1$) we see that for any face $S$ and any smearing vector field $f$  and for any function $\Psi'\in\cyl$
\[
\int_{\gcon}X_{S,f}(\Psi')\,d\mu=0.
\]
As it was shown in \cite{Okolow:2003pk} the only measure satisfying the above condition coincides with the measure defined on $\gcon$ by the state $\omega_0$ described in the example.

In conclusion,
\begin{equation}
\omega\ =\ \omega_0.
\end{equation}
\end{proof}
\section{Closing remarks}
\label{concl}
As we have emphasized the uniqueness result proved in the last
section is reassuring for the LQG program, and it shows that
diffeomorphism invariance can sometimes be a powerful remedy
against complications that one expects based on what one knows
about background dependent  field theories. Our result is based on
certain, albeit reasonable, assumptions, therefore an immediate
question is whether it can be generalized.

Certainly the result holds for any enlargement of the symmetry
group that contains the diffeomorphisms considered above. Whether
it also holds for smaller extensions, or even for only the
analytic diffeomorphisms is an open question. We feel however that
as soon as the subgroup  of diffeomorphisms  is big enough to
contain 'local ones', application of the techniques used above
should be straightforward.

Also, even if uniqueness were to break down if only invariance
under analytic automorphisms is required, it is not clear how
relevant the result would be physically, as it would heavily
involve details of the structure of analytic diffeomorphisms on a
given manifold $\Sigma$.

Another way to generalize the result would be to consider, instead
of the flux operators, the unitary groups that they generate, and
ask for diffeomorphism invariant representations in which these
groups are strongly continuous. In \cite{Sahlmann:2003in} this
setting was considered, however the results were not satisfactory
due to the more complicated domain questions arising.
A more satisfying result was recently obtained by Fleischhack in
\cite{christ2}. It does not make the assumption of a common dense
domain for all flux operators and cylindrical functions
that is implicit in  our treatment. However, it needs an additional
assumption on the action of the bundle automorphisms in the
representation.

Finally, as for background dependent theories, at least the definition of the
kinematical algebra $\skripta$ applies in principle. Whether one
expects the type of variables used, to be well defined in the quantum
theory is certainly a difficult question.
Still it seems worthwhile to look for non-diffeomorphism invariant
representations of $\skripta$ and see if they can be put to use in
physics.

Another interesting starting point for future work is the observation
that the uniqueness theorem fails if a rather
innocent looking assumption -- that of compact support on $S$ for the
smearing functions $f$ used in the flux variables $E_{S,f}$ -- is removed:
Consider the example $\Sigma=\R^2$, $G=$U(1),
and drop the assumption of compact support for the smearing functions.
The hyper-surfaces $S$ are one dimensional in this case.
Let us also choose an orientation for $\Sigma$. From that orientation,
together with the orientation on the normal bundle of a given $S$ we
can equip $S$ with an intrinsic orientation, and thus integrate one-forms
on $S$. Then we can define
\begin{align*}
\omega(\hat \Psi)&:=\omega_0(\hat \Psi),\\
\omega(\hat \Psi\hat X_{S_1,f_1}\ldots\hat X_{S_n,f_n})
&:= \int_{S_1}df_1\ldots \int_{S_n}df_n\,\omega_0(\hat \Psi).
\end{align*}
It is easy to check that $\omega$ defines a state on $\skripta$ and is
manifestly invariant under the action of orientation-preserving
diffeomorphisms. Obviously it
is different from $\omega_0$ and thus would constitute a
counterexample to our uniqueness result, were it not for the fact that
for smearing functions $f$ with compact support in $S$,
$\int_Sdf$=0. Hence under the assumptions made in this paper,
$\omega=\omega_0$, and there is no contradiction.

Since obvious generalizations of this state to a higher dimensional
situation seem to fail, the existence of $\omega$ for smearing
functions without compact support might just be a peculiarity of
$D=2$.
However, just as for $\omega$ the endpoints of the lines $S$ can be
used to ``anchor'' diffeomorphism invariant information in the state,
it is not inconceivable that similarly points of the boundary of
hyper-surfaces $S$ in which the boundary has a
lower differentiability than $C^m$ might be used to that end in higher
dimensions.\footnote{In a similar way, one might intuitively
  understand the need to use semi-analytic smearing functions, not
  just, say, continuous ones: Singular points (from the semianalytic
  perspective) of the smearing functions could not be removed by
  semianalytic diffeomorphisms and evaluation of the function at such points
  would thus constitute diffeomorphism invariant data that could give
  rise to other diffeomorphism-invariant states.}
In any case, the restriction to compact support does not seem to be
unphysical.\footnote{And, as for applications to Loop Quantum Gravity,
all results that use smearing functions with non-compact support
(such as the definition of the volume operator) can be recovered by taking
appropriate limits once the state is fixed to be $\omega_0$.}
It can be viewed as analogous to the smoothness and decay properties
assumed for smearing functions in standard quantum field theory.
A more detailed investigation into these issues will
be carried out elsewhere.

A useful for LQG outcome of our work is introducing the semianalytic
category.
The corresponding diffeomorphisms form
a subgroup of the C$^{m}$
diffeomorphism group, the group of symmetries
induced by the action of the diffeomorphisms of $\Sigma$ in  the
classical phase-space and preserving our classical
Ashtekar-Corichi-Zapata algebra. The relevance of this symmetry
group consists in its local character (as opposed to the analytic
diffeomorphisms). For example, the symmetry group  provides a new,
elegant version of a map $\cyl\rightarrow \cyl^*$ (the algebraic
dual) which averages with respect to the (allowed) diffeomorphisms
of $\Sigma$. This application has been recently implemented in
\cite{alrev}\footnote{Except that our definition of the extension
of the analytic diffeomorphism group has changed since then}
(see also  \cite{almmt,lm,ttrev}). There are several  similar,
non-equivalent extension of the analytic diffeomorphisms recently
introduced in literature.  One of them is due to  Fleischhack \cite{christ2}
who also applies  the theory of the stratifications.
Another one  was considered by Rovelli and Fairbairn \cite{rovelli}.
They even advocate  the relevance of non-differentiable homeomorphisms
in the classical Einstein's Gravity. The Rovelli--Fairbairn generalized
diffeomorphisms, however, are defined to be smooth everywhere
except a finite set of points, therefore they would not be useful
in our case.
\\
\\
{\bfseries \Large Acknowledgments}
\\
We thank Abhay Ashtekar, Klaus
Fredenhagen, Detlev Buchholz, Stefan Hollands and Bob Wald for urging us to
investigate the uniqueness issue. Stanis{\l}aw Woronowicz gave us
a useful technical suggestion we appreciate very much. We thank
Christian Fleischhack for drawing our attention to the theory of the
semianalytic sets.
JL and AO were partially supported by Polish KBN grants: 2 P03B
06823 and 2 P03B 12724. TT was supported in part by a grant from
NSERC of Canada to the Perimeter Institute for Theoretical Physics.
Parts of this work were supported by NSF grant PHY-00-90091 and the
Eberly research funds of Penn State.
\\
\\
\appendix
\section{semianalytic category, details}
\subsection{semianalytic functions in $\R^n$}
In this section we introduce semianalytic functions, semianalytic
manifolds and semianalytic geometry. We will take advantage of the
results of the theory of semianalytic sets
\cite{loj,bier-pier}\footnote{We thank Christian Fleischhack
\cite{chr} for drawing our attention to the theory
 of semianalytic sets.}

Briefly speaking, a real valued function $f$ defined on an open
subset of ${\cal U}\subset\R^n$ will be called semianalytic if it is
analytic on an open and dense subset of ${\cal U}$, and if the
non-analyticity surfaces have also an appropriate analytic
structure, and if the restrictions of $f$ to the non-analyticity
surfaces are again analytic in an appropriate sense. To introduce
our definition, we need a notion of a semianalytic partition of
${\cal U}$.   Consider in ${\cal U}$ a {\it finite} sequence of
equalities and/or inequalities, namely
\begin{align}
h_1(x)\ &\sigma_1\ 0,\nonumber\\
&\ldots \label{ineq}\\
 h_N(x)\ &\sigma_N\ 0\,,\nonumber
\end{align}
where each $\sigma_i$ is either of the three relations $>\, ,\, <\,
,\,=$, and $\{h_1,\, ...\, ,h_N\}$ is a set of analytic functions
defined on a domain containing  ${\cal U}$. More formally, there is
defined a map
\begin{equation}
\sigma\ :\ h=\{h_1,...,h_N\}\ \rightarrow\ \{>,=,<\}\label{sigma}
\end{equation}
and  in (\ref{ineq}) we denoted
\begin{equation}
\sigma_I\ :=\ \sigma(h_I),
\end{equation}
where the integer $I$ runs from 1 to $N$. The set of the conditions
(\ref{ineq}) determines the following subset of ${\cal U}$,
\begin{equation}
{\cal U}_{h,\sigma}\ =\ \{x\in{\cal U}\,:\, (\ref{ineq})\}.
\label{Uhsig}
\end{equation}

\begin{defi}\label{part}
Given a finite set $h$ of real valued analytic functions defined on
a neighborhood of an open subset $\U$ of $\R^n$, the corresponding
semianalytic partition of $\U$ is the set of all the subsets ${\cal
U}_{h,\sigma}\subset\U$ defined by (\ref{ineq},\ref{Uhsig}) such
that $\sigma$ is an arbitrary map (\ref{sigma}). Given ${\cal U}$
and $h$ as above, the partition will be denoted by ${\cal P}({\cal
U},h)$.
\end{defi}

Obviously, every semianalytic partition covers $\U$
\begin{equation}
{\cal U}\ =\ \bigcup_{\sigma}{\cal U}_{h,\sigma}
\end{equation}
 where $\sigma$ runs through all the maps (\ref{sigma}). Also,
\begin{equation}
\sigma\not=\sigma'\ \ \ \ \Rightarrow\ \ \ \ {\cal U}_{h,\sigma}\cap
{\cal U}_{h,\sigma'}=\emptyset\, ,
\end{equation}
and a set ${\cal U}_{h,\sigma}$ may be empty itself. Another obvious
property is that given a semianalytic covering ${\cal P}({\cal
U},h)$ and an open subset ${\cal V}\subset {\cal U}$, the family $h$
of functions defines a semianalytic covering ${\cal P}({\cal V},h)$.

Now, we are in a position to define a semianalytic function:
\begin{defi}\label{semianalytic}
A function $f:{\cal U}\rightarrow\R^m$, where ${\cal U}$ is an open
subset of $\R^n$, is called semianalytic if  every $x\,\in\,{\cal
U}$ has an open neighborhood $\tilde{\cal U}$ equipped with a
semianalytic partition ${\cal P}(\tilde{\cal U},h)$, such that for
every $\tilde{\cal U}_{h,\sigma}\in {\cal P}(\tilde{\cal U},h)$
there is an analytic function $f_\sigma:\tilde{{\cal
U}}\rightarrow\R^m$, such that
\begin{equation}
f\vert_{\tilde{\cal U}_{h,\sigma}}\ =\ f_\sigma\vert_{\tilde{\cal
U}_{h,\sigma}},\label{restr}
\end{equation}
that is, such that $f_\sigma$ coincides with $f$ on ${\tilde{\cal
U}_{h,\sigma}}$.
\end{defi}
Given a semianalytic function $f$ and a point $x$ in its domain, a
semianalytic partition ${\cal P}(\tilde{\cal U},h)$ which has the
properties described in Definition \ref{semianalytic} will be called
{\it compatible} with $f$ at the point $x$. There are infinitely
many semianalytic partitions compatible with a given $f$ at $x$.
Clearly, if $f:{\cal U}\rightarrow \R^n$ is semianalytic, and ${\cal
V}\subset {\cal U}$ is open, then the restriction function
$f_{|{\cal V}}$ is semianalytic. Given a semianalytic covering
${\cal P}(\tilde{\cal U},h)$ compatible with $f$, and an open subset
$\tilde{\cal V}\subset\tilde{\cal U}\cap {\cal V}$, the semianalytic
covering ${\cal P}(\tilde{\cal V},h)$ is compatible with $f_{|{\cal
V}}$.
\medskip

\noindent{\bf Example} Consider a function $f:\R\rightarrow \R$
analytic on every closed interval $[n,n+1]$. $f$ is semianalytic. A
semianalytic partition ${\cal P}(\tilde{\U},h)$ compatible with $f$
at $x_0$ is defined for the open interval
\begin{equation}
\tilde{\U}\ :=\ \big]\,[x_0]-1,\,[x_0]+1\,\big[
\end{equation}
(we denote by $\big] a,b\big[$ the open interval bounded by
$a,b\in\R$ and by $[a]$ the integer part of $a$.)
 by the set $\{h_{-1},h_{0}, h_{1}\}$  of functions
\begin{equation}
h_{-1}(x)\ =\ x-[x_0]+1,\ \ h_0(x)\ =\  x-[x_0],\ \ h_1(x)\ =\
 x-[x_0]-1.
\end{equation}
\medskip

\begin{prop} \label{ff} Let $f_1:{\cal U}\rightarrow\R$, and
$f_2:{\cal U}\rightarrow\R^m$ be two semianalytic functions where
${\cal U}$ is an open subset of $\R^n$. Then the functions
\begin{equation}
{\cal U}\ni x \mapsto f_1(x)f_2(x)\in\R^m,\ \ \ \ \ {\cal U}\ni x
\mapsto (f_1(x),f_2(x))\in\R^{m+1},
\end{equation}
are also semianalytic.
\end{prop}
\begin{proof}
Let $x\in{\cal U}$. Let ${\cal P}(\tilde{\cal U}^{(1)},{h}^{(1)})$
be a semianalytic  partition compatible with $f_1$ at $x$, and
  ${\cal P}(\tilde{\cal U}^{(2)},{h}^{(2)})$
be a semianalytic  partition compatible with $f_2$ at $x$. The proof
becomes obvious if  we construct  a single semianalytic partition
compatible at $x$ with both the functions.
 The natural choice is just the semianalytic partition of the
intersection
\begin{equation}
\tilde{\cal U}\ :=\ \tilde{\cal U}^{(1)}\cap  \tilde{\cal U}^{(2)}
\end{equation}
defined by the set of functions
\begin{equation}
{h}\ :=\ h^{(1)}\cup h^{(2)}.
\end{equation}
Indeed, it is enough to notice, that for every $\tilde{\cal
U}_{h,\sigma}\in{\cal P}({\cal U},h)$ there are some $\tilde{\cal
U}^{(1)}_{h^{(1)},\sigma^{(1)}}\in {\cal P}(\tilde{\cal
U}^{(1)},{h}^{(1)})$ and
 $\tilde{\cal U}^{(2)}_{h^{(2)},\sigma^{(2)}}\in
{\cal P}(\tilde{\cal U}^{(2)},{h}^{(2)})$ such that
\begin{equation}
\tilde{\cal U}_{h,\sigma}\ \subset \tilde{\cal U}^{(1)}_{h^{(1)},
\sigma^{(1)}}\ \ \ \ {\rm and}\ \ \ \ \tilde{\cal U}_{h,\sigma}\
\subset \tilde{\cal U}^{(2)}_{h^{(2)}, \sigma^{(2)}}.
\end{equation}
\end{proof}

It is obvious, that if $f:\U\rightarrow\R$ is a  semianalytic
function and it does not vanish on an open set $\U$, then
$\frac{1}{f}$ is also semianalytic. This fact will be important in
construction of semianalytic partitions of unity. They are useful
owing to Proposition \ref{ff}.

We turn now, to the issue of the morphisms of the semianalytic
functions. It is obvious, that  every analytic map $\phi: {\cal U}\
\rightarrow\ {\cal U}'$ between two open subsets ${\cal U}\subset
\R^n$ and ${\cal U}'\subset \R^{n'}$ pullbacks all the semianalytic
functions defined on ${\cal U}'$ into semianalytic functions defined
on ${\cal U}$. The following proposition shows, that the same is
true for a semianalytic map.
\begin{prop}\label{fphi}
Let ${\cal U}\subset \R^n$ and ${\cal U}'\subset \R^{n'}$ be open
subsets. Suppose the functions $f':{\cal U}'\rightarrow \R^m$ and
$\phi:{\cal U}\rightarrow  {\cal U}'$ are semianalytic. Then, the
composition function\\ $f'\circ \phi:{\cal U}\rightarrow \R^m$ is
semianalytic.
\end{prop}
\begin{proof}
The idea of the proof is simple: we construct a suitable partition
of $\U$ using the inverse image of a given partition of $\U'$
compatible with $f'$. The inverse image is not, in general,
semianalytic, but we show it can be sub-divided into a semianalytic
partition.

\begin{lemm} Let ${\cal P}(\tilde{\U}',h')$ be a semianalytic
partition. Let $\phi:\U\rightarrow\tilde{\U}'$ be a semianalytic
function, where the subset  $\U\subset \R^{n}$ is open. For every
$x_0\in\U$ there exists an open neighborhood $\tilde{\U}$ and a
semianalytic partition ${\cal P}(\tilde{\U},\tilde{h})$ such that
for every element of ${\cal P}(\tilde{\U},\tilde{h})$, say
$\tilde{\U}_{\tilde{h},\tilde{\sigma}}$, there is an element of
  ${\cal P}(\tilde{\U}',h')$, say $\tilde{\U}'_{h',\sigma'}$, such that
\begin{equation}
\tilde{\U}_{\tilde{h},\tilde{\sigma}}\subset
\phi^{-1}\left(\tilde{\U}'_{h',\sigma'}\right).
\end{equation}
\end{lemm}
\begin{proof}
Let $x_0\in\U$. We will construct a partition ${\cal
P}(\tilde{\U},\tilde{h})$ which satisfies the conclusion. If $\phi$
is analytic, then the set of the pullbacks of all the functions
$h'_I\in h'$ defines a suitable partition of the whole $\U$. In
general, $\phi$ is not analytic. However, it gives rise to a family
of analytic functions $\phi_\sigma$ defined in some neighborhood
$\tilde{\U}$ of $x_0$ via Definition \ref{semianalytic} (with $f$
being replaced by $\phi$). We choose $\tilde{\U}$ small enough, such
that all the images $\phi_\sigma(\tilde{\U})$ are contained in the
given $\tilde{\U}'$). Hence, consider a semianalytic partition
${\cal P}(\tilde{\U},h)$ compatible with $\phi$ at $x_0$, and the
corresponding family of analytic functions
$\phi_\sigma:\tilde{\U}\rightarrow \tilde{\U}'$. We define
$\tilde{h}$ to be the set of functions formed by $(i)$ all the
functions $h'_I\circ \phi_\sigma$ defined by all the functions
$\phi_\sigma$ and all $h'_I\in h'$, and $(ii)$ all the functions
${h}_I\in h$. Let us demonstrate that the corresponding semianalytic
partition ${\cal P}(\tilde{\U},\tilde{h})$ satisfies the conclusion.
Let $\tilde{\sigma}:\tilde{h} \rightarrow \{>,\, =,\, <\}$ be an
arbitrary map. Denote
\begin{align}
\sigma\ &:=\ \tilde{\sigma}\vert_h \label{1}\\
\sigma'\ &:=\ \tilde{\sigma}\vert_{\phi^*_\sigma(h')}\label{2}
\end{align}
where $\sigma$ in the second line is the one introduced in the first
line, and $\phi^*_\sigma(h')$ is the set of pullbacks of the
elements of $h'$ by using $\phi^*_\sigma$. Now, it follows directly
from (\ref{1})  (see (\ref{ineq}) with $h_I$ and $\sigma_I$ being
themselves as well as being replaced by $\tilde{h}_I$ and
$\tilde{\sigma}_I$) that
\begin{equation}
\tilde{\U}_{\tilde{h},\tilde{\sigma}}\subset
\tilde{\U}_{{h},{\sigma}}.
\end{equation}
 On the other hand,  the second line (\ref{2}) means that
\begin{equation}
\phi_\sigma(\tilde{\U}_{\tilde{h},\tilde{\sigma}})\ \subset\
\tilde{\U}'_{h',\sigma'}.
\end{equation}
The combination of the last two facts with
\begin{equation}
\phi_\sigma\vert_{\tilde{\U}_{{h},{\sigma}}}\ =\
\phi\vert_{\tilde{\U}_{{h},{\sigma}}}\
\end{equation}
concludes the proof of Lemma.
\end{proof}
We go back to the proof of Proposition.  Given $x_0\in \U$, consider
the point $\phi(x_0)$ and a partition ${\cal P}(\tilde{\U}',h')$
compatible with the function $f'$ at $\phi(x_0)$. Let ${\cal
P}(\tilde{\U},\tilde{h})$ be a partition provided by Lemma. For
every $\tilde{\U}_{\tilde{h},\tilde{\sigma}}\in {\cal
P}(\tilde{\U},\tilde{h})$ use the pair $\sigma,\sigma'$ defined by
(\ref{1},\ref{2}). The function $f'_{\sigma'}\circ\phi_\sigma$ is
the wanted analytic extension of $f'\circ \phi\vert_{
\tilde{\U}_{\tilde{h},\tilde{\sigma}}}$.
\end{proof}

In general, the inverse of an invertible semianalytic function is
not necessarily semianalytic.  However, a carefully formulated set
of assumptions ensures the semianalyticity of the inverse.

\begin{prop}\label{inverse}
Let  $\phi:\U\rightarrow \U'$ be a semianalytic and bijective
function, where $\U,\, \U'\subset \R^n$ are  open.  Suppose, that
for every $x_0\in \U$ there exists a semianalytic partition ${\cal
P}(\tilde{\U},h)$ compatible with $\phi$ at $x_0$, and such that for
every $\tilde{\U}_{h,\sigma}\in {\cal P}(\tilde{\U},h)$ the
restriction $\phi\vert_{\tilde{\U}_{h,\sigma}}$ is extendable to an
analytic, injective function $\phi_\sigma:\tilde{\U}\rightarrow
\U'$, such that: $(i)$ $\phi_\sigma(\tilde{\U})$ is an open subset
of $\R^n$, and $(ii)$ the inverse
$\phi_\sigma^{-1}:\phi_\sigma(\tilde{\U})\rightarrow\tilde{\U}$ is
analytic. Then, $\phi^{-1}$ is semianalytic.
\end{prop}
\begin{proof}
Given a point $x'_0\in \U'$, let ${\cal P}(\tilde{\U},h)$ be a
partition compatible with $\phi$ at $x_0=\phi^{-1}(x'_0)\in \U$.
Suppose ${\cal P}(\tilde{\U},h)$ and $\phi$  satisfy the
assumptions. We have to construct a semianalytic partition of a
neighborhood $\tilde{\U}'$ of $x'_0$ compatible with $\phi^{-1}$.
We choose $\tilde{\U}'$ such that all the inverse functions
$\phi_\sigma^{-1}$ are well defined, namely
\begin{equation}
\tilde{\U}'\ =\ \bigcap_\sigma\phi_\sigma(\tilde{\U}).
\end{equation}
Mapping with $\phi$ the partition ${\cal P}(\tilde{\U},h)$ we get a
partition of $\tilde{\U}'$ which consists of the sets
\begin{equation}
\phi\left(\tilde{\U}_{h,\sigma}\right)\cap\tilde{\U}',
\end{equation}
given by all the elements $\tilde{\U}_{h,\sigma}\in {\cal
P}(\tilde{\U},h)$ . For every set
$\phi\left(\tilde{\U}_{h,\sigma}\right)\cap\tilde{\U}'$ we have
\begin{equation}\label{ext}
\phi^{-1}\vert_{\phi\left(\tilde{\U}_{h,\sigma}\right)\cap\tilde{\U}'}
\ =\ \phi^{-1}_\sigma\vert_{
\phi\left(\tilde{\U}_{h,\sigma}\right)\cap\tilde{\U}'}
\end{equation}
where $\phi_\sigma^{-1}$ is the analytic function provided by the
assumptions. That would be sufficient for the semianalyticity of
$\phi^{-1}$ if the constructed partition were semianalytic. We do
not know if it is the case, though. However, we will subdivide the
partition in such a way, that the result is a semianalytic partition
without any doubt. Establishing that refined partition will be
enough to complete the proof by referring to (\ref{ext}). The needed
semianalytic partition is defined in the following way. First, we
fix a subset $\phi\left(\tilde{\U}_{h,\sigma}\right)\cap\tilde{\U}'$
and use the corresponding analytic function $\phi^{-1}_{\sigma}$ to
pullback all the functions $h_I\in h$ from $\tilde{\U}$ onto
$\tilde{\U}'$. Denote the resulting set of analytic, real valued
functions defined on $\tilde{\U}'$ by ${\phi_{\sigma}^{-1}}^*h$, and
consider the corresponding semianalytic partition ${\cal
P}(\tilde{\U}',{\phi_\sigma^{-1}}^*h)$. It is easy to see, that
\begin{equation}
\phi\left(\tilde{\U}_{h,\sigma}\right)\cap\tilde{\U}' \in {\cal
P}(\tilde{\U}',{\phi^{-1}_{\sigma}}^*h).
\end{equation}
Next,   enlarge the set ${\phi^{-1}_{\sigma}}^*h$ corresponding to a
given $\sigma$ by taking the union with respect to all the $\sigma$s
(\ref{sigma}),
\begin{equation}
h'\ =\ \bigcup_\sigma {\phi^{-1}_{\sigma}}^*h.
\end{equation}
Consider the  semianalytic partition ${\cal P}(\tilde{\U}',h')$
defined by $h'$. This partition just divides every
$\phi\left(\tilde{\U}_{h,\sigma}\right)\cap\tilde{\U}'$ into smaller
subsets of $\tilde{\U}'$, that is it consists of subsets of the sets
$\phi\left(\tilde{\U}_{h,\sigma}\right)\cap\tilde{\U}'$. This
concludes the proof.
\end{proof}

\begin{coro}
\label{inverse2} Suppose $\phi:\U\rightarrow\U'$ is a
diffeomorphisms of the differentiability class $C^{m}$, where
$\U,\U'\subset\R^n$ are open and $m>0$. If $\phi$ is semianalytic,
then so is $\phi^{-1}: \U'\rightarrow\U$.
\end{coro}

\begin{proof}
 Let us assume that $\phi$ satisfies the assumptions made in
  Corollary \ref{inverse2} and consider an arbitrary point $x_0$ in the
  domain $\cal U$. Since $\phi$ is semianalytic, we can find: (a) a
  neighborhood $\tilde{\cal U}$ of $x_0$, (b) a semianalytic
  partition
  ${\cal P}(\tilde{U},h)$, and (c) for
  every $\tilde{U}_{h,\sigma}\in {\cal P}(\tilde{U},h)$ an analytic
  function $\phi_\sigma$ defined on $\tilde{U}$, which
   coincides with $\phi$ on $\tilde{U}(h,\sigma)$.

It would be sufficient to show that the data (a)-(c) can be chosen
in such a way that every function $\phi_\sigma$ of (c) has a non
degenerate derivative $D\phi_\sigma$ at every point of $\tilde{\cal
U}$. Then the hypothesis of Proposition \ref{inverse} would be
satisfied. Certainly the derivative of $\phi$  is nowhere degenerate
in $\cal U$. Therefore, for every function $\phi_\sigma$ of (c)
there is an open subset of $\tilde{\cal U}$ such that the derivative
of $\phi_\sigma$ is non-degenerate. The problem is that the subsets
of points on which the derivatives are non-degenerate may be too
small. They may be smaller than $\tilde{\cal U}$, and some of them
may even not contain the point $x_0$ at all. Therefore the data
(a)-(c) is not yet sufficient to apply Prop. \ref{inverse}.

We therefore define new data (a')-(c') given by shrinking the
neighborhood $\tilde{\cal U}$  appropriately. For every
$\tilde{U}_{h,\sigma}\in{\cal P}(\tilde{U},h)$ consider the subset
$S_\sigma\subset\tilde{U}$ of points  such that the function
$\phi_\sigma$ has a nondegenerate derivative. Note, that $S_\sigma$
contains the completion $\overline{\tilde{U}_{h,\sigma}}$. Indeed,
it follows from the continuity of $D\phi$ and $D\phi_\sigma$. As a
new $\tilde{\cal U}'$ we take the intersection of all the subsets
$S_\sigma$ such that $x_0\in\overline{\tilde{U}_{h,\sigma}}$.
$\tilde{\cal{U}}'$ then constitutes new data (a'). A new partition
(b') and functions (c') are given just by restricting the previous
(b),(c) to $\tilde{\cal U}'$. (a')-(c') then fulfill the assumptions
of Prop. \ref{inverse}.
\end{proof}
Once we have generalized the notion of analytic structure into the
notion of the semianalytic structure, it is natural to introduce new
partitions by relaxing in Definition \ref{part} the assumption that
the functions constituting the set $h$ are analytic, and replace it
by a condition that they be semianalytic. Let us do it,  apply the
same notation as in Definition \ref{part} to a finite set of {\it
semi}analytic functions $h$ and call the result a semi-semianalytic
partition.

Given any  partition of a set into subsets, another  partition is
called finer if every element of the first partition is a finite
union of elements of the second partition.

\begin{lemm}\label{normpart} Suppose ${\cal P}(\U,h)$ is a
semi-semianalytic partition of an open $\U\subset\R^n$. Then, every
$x\in\U$ has a neighborhood $\tilde{\cal U}$ which admits a
semianalytic partition finer than ${\cal P}(\tilde{\cal U},h)$.
\end{lemm}

\begin{proof} Let $x_0\in \U$.  There is a neighborhood $\tilde{\U}$
of $x_0$ which admits a semianalytic partition  ${\cal
P}(\tilde{\U},f)$ compatible with all the (semianalytic functions)
elements of $h$. As before, we start with collecting all the
analytic functions available. Firstly, all the elements $f_I\in f$
are  analytic functions defined on $\tilde{\U}$. Secondly, for every
assignment $\sigma:f\rightarrow \{>,\,=,\,<\}$, every element
$h_I\in h$ defines an analytic function $h_{I\sigma}$. Given
$\sigma$ denote the set of the functions $h_{I\sigma}$ such that
$h_I\in h$ is arbitrary, by $h_\sigma$. The resulting set of the
analytic functions is
\begin{equation}
\tilde{h}\ :=\ f\cup\bigcup_\sigma h_\sigma.
\end{equation}
Our candidate for a semianalytic partition of $\tilde{\U}$
compatible with ${\cal P}(\U,h)$ is the semianalytic partition
defined by the set of functions $\tilde{h}$. Consider an arbitrary
\begin{equation}
\tilde{\sigma}:\tilde{h}\rightarrow \{>,\,=,\,<\},
\end{equation}
and the corresponding set $\tilde{\U}_{\tilde{h},\tilde{\sigma}}\in
{\cal P}(\tilde{\U},\tilde{h})$. We have to point out an element
$\U_{h,\sigma'}$ which contains
$\tilde{\U}_{\tilde{h},\tilde{\sigma}}$. It is defined as follows.
Consider
\begin{equation}
\sigma\ :=\ \tilde{\sigma}\vert_f.
\end{equation}
Using this $\sigma$ select another subset of $\tilde{h}$, namely
$h_\sigma$. The restriction
\begin{equation}
\tilde{\sigma}\vert_{h_\sigma},
\end{equation}
defines naturally an assignment $\sigma':h\rightarrow
\{>,\,=,\,<\}$, namely
\begin{equation}
\sigma'(h_I)\ :=\ \tilde{\sigma}(h_{I\sigma}).
\end{equation}
It is easy to check that
$\tilde{\U}_{\tilde{h},\tilde{\sigma}}\subset \U_{h,\sigma'}$.
\end{proof}

Finally, our interest in the semi-analytic sets  is a consequence of
a certain strong result of that theory (\cite{loj}, see Proposition
2.10 in \cite{bier-pier}) which we translate now into the terms of
the semianalytic partitions.

We call a semianalytic partition analytic partition if every element
of the partition is a connected, analytic submanifold.

The result we are referring to reads:
\begin{prop}
For every semianalytic partition ${\cal P}({\cal U},h)$ of an open
$\U\subset\R^n$, every point $x\in\U$ has a neighborhood
$\tilde{\cal U}$ which admits an analytic partition finer than
${\cal P}(\tilde{\cal U},h)$.
\end{prop}

\subsection{Semianalytic manifolds and submanifolds}
In this subsection, $\Sigma$ is an $n$ dimensional differential
manifold. Henceforth we will be assuming that $\Sigma$ and all the
considered functions are of a differentiability class C$^{m}$, where
$m>0$.

By analogy with the definitions of an analytic structure, analytic
function, and analytic submanifold, we introduce now natural
semianalytic generalizations. The generalization is possible due to
Propositions \ref{ff}, \ref{fphi}, \ref{inverse} of the previous
subsection.

We denote below  an  atlas of $\Sigma$ by
$\{(\U_I,\chi_I)\}_{I\in{\cal I}}$ , where ${\cal I}$ is some
labeling set, $\{\U_I\}_{I\in{\cal I}}$ is an open covering of
$\Sigma$, and $\{\chi_I\}_{I\in{\cal I}}$ is a family of
diffeomorphisms $\chi_I:\U_I\rightarrow \U'_I\subset\R^n$.

\begin{defi}
An atlas $\{(\U_I,\chi_I)\}_{I\in{\cal I}}$ of $\Sigma$ is called
semianalytic if for every pair $I,J\in{\cal I}$ the map
\begin{equation}
\chi_J\circ\chi_I^{-1}:\ \chi_I(\U_I\cap\U_J)\ \rightarrow\
\chi_J(\U_I\cap\U_J)
\end{equation}
is semianalytic. The diffeomorphisms $\chi_I$ are called semianalytic charts.\\
 A semianalytic structure on
$\Sigma$ is a maximal semianalytic atlas. A semianalytic manifold is
a differential manifold endowed with a semianalytic structure.
\end{defi}
\medskip

\begin{defi}
 Given two semianalytic manifolds $\Sigma$ and $\Sigma'$,
 a map $f:\Sigma\rightarrow \Sigma'$ is called semianalytic
 if for every  semianalytic chart $\chi_I$ of $\Sigma$,
 and every semianalytic chart $\chi'_{I'}$ of $\Sigma'$
 the function $\chi'_{I'}\circ f\circ\chi_I^{-1}$ (whenever
 the composition can be applied) is semianalytic.
\end{defi}

In particular, if
\begin{equation}
\Sigma'\ =\ \R^{n'}
\end{equation}
and the semianalytic structure is the natural one defined by the
atlas $\{(\R^{n'},\, {\rm id})\}$, then the map $f$ is a
semianalytic function defined on $\Sigma$.

\begin{defi}
A semianalytic submanifold of a semianalytic manifold $\Sigma$ is a
subset $S\subset \Sigma$ such that for every $x\in S$, there is a
semianalytic chart $\chi_I$ defined in a neighborhood $\U_I$ of $x$,
such that
\begin{equation}\label{subm}
\begin{split}
\chi_I(S\cap \U_I)\ =\ \{(x^1,\,...\,,x^n)\in\R^n\ :\
x^1=...=x^{n-n'}=0,\\
\qquad 0<x^{n-n'+1}<1,\,...\,,\,0<x^{n}<1\},
\end{split}
\end{equation}
where $n'$ is a non-negative integer, $n'\le n$, and $n'$ is called
the dimension of $S$.
\end{defi}

\begin{defi}
An $n'$ dimensional  semianalytic submanifold with boundary of
$\Sigma$ is a subset $S\subset \Sigma$ such that for every $x\in S$,
there is a semianalytic chart $\chi_I$ defined in a neighborhood
$\U_I$ of $x$, such that either (\ref{subm}) or
\begin{equation}\label{submb}
\begin{split}
\chi_I(S\cap \U_I)\ =\ \{(x^1,...,x^n)\in\R^n\ :\
x^1=...=x^{n-n'}=0,\\
0\le x^{n-n'+1} < 1,\, 0 < x^{n-n'+2}<1,\, ...\,,\,0< x^{n}<1\}
\end{split}.
\end{equation}
\end{defi}

The key property of the semianalytic submanifolds crucial in our
work is:

\begin{prop}\label{finiteness}
Let $S_1$ and $S_2$ be two semianalytic submanifolds of a
semianalytic manifold $\Sigma$. Suppose $x\in S_1\cap S_2$. Then,
there is an open neighborhood ${\cal W}$ of $x$ in $\Sigma$, such
that ${\cal W}\cap S_1\cap S_2$ is a finite, disjoint union of
connected semianalytic submanifolds.
\end{prop}
\noindent{\bf Remark.} What is crucial for us in the conclusion of
Proposition \ref{finiteness} is the {\it finiteness} of the
partition and the {\it connectedness} of its elements. After all, an
infinite set of disjoint, embedded intervals may also form a single
submanifold, disconnected though. Those two properties
simultaneously hold due to the (semi)analyticity.
\begin{proof}
For every point $x\in S_1\cap S_2$, there is a neighborhood ${\cal
W}$ which can be mapped by a semianalytic chart into an open subset
${\U}\subset\R^n$. The intersection ${\cal W}\cap S_1\cap S_2$ is
mapped into a subset of $\U$ described by a finite family of
equalities of the form (\ref{ineq}) defined by some fixed family of
semianalytic functions $h_I$ and  relations $\sigma_I\ =\ `='$ (the
definition of a semianalytic submanifold involves also inequalities,
however ${\cal W}$ can be chosen such that the latter ones are
satisfied at every point in ${\cal W}$; we are assuming this is the
case). Hence, the intersection is an element of the
semi-semianalytic partition defined by the family of the
semianalytic functions $h_I$. Due to Lemma \ref{normpart}, if we
choose the neighborhood ${\cal W}$ of the point $x$ appropriately,
then the intersection ${\cal W}\cap S_1\cap S_2$ is a finite union
of elements of certain semianalytic partition. Finally, via the
result quoted in the previous subsection, the neighborhood $\cal W$
can be chosen such that every element of a semianalytic partition of
the image ${\cal U}$ is a finite, disjoint union of connected
analytic submanifolds. Their inverse image by the chart defines the
decomposition of the intersection ${\cal W}\cap S_1\cap S_2$ into
semianalytic submanifolds.
\end{proof}

In the paper we are using extensively two particular classes of
submanifolds: edges and faces.

\begin{defi}
A semianalytic edge is a connected, 1-dimensional semianalytic
submanifold of $\Sigma$ with 2-point boundary.
\end{defi}

\begin{defi}\label{face}
A face is a connected, codimension 1 semianalytic submanifold of
$\Sigma$ whose normal bundle is equipped with an orientation.
\end{defi}

The property of the semianalytic structures which distinguishes them
so much from the analytic ones is local character of the spaces of
the semianalytic functions and semianalytic diffeomorphisms. That
feature is guaranteed by the existence of a partition of unity
compatible with an arbitrary open covering. We formulate this fact
precisely now, in the form we refer to in the proof of our main
theorem:

\begin{prop}
Suppose $\mc{W}\subset\Sigma$ is a compact subset. Let
$\mc{U}_{I}\subset\Sigma$, $I=1,\,\ldots,\,N$, be a family of open
sets which covers $\mc{W}$. There exists a family of C$^{m}$
semianalytic functions $\phi_{I}:\Sigma\rightarrow\R$,
$I=1,\,\ldots\,,N$ such that for every $I$,
\begin{equation}
\supp \phi_I\ \subset\ \mc{U}_{I}
\end{equation}
and
\begin{equation}
\sum_I\phi_I\vert_{\mc{W}}\ =\ 1.
\end{equation}
\end{prop}
\begin{proof}
The proof is standard owing  to the following two properties of the
semianalytic functions:

$(i)$ For every open  ball in $\R^D$, there is a C$^{m}$
semianalytic function greater than zero at every point inside the
ball and identically zero everywhere else.

$(ii)$ If $f$ is a nowhere vanishing C$^{m}$ semianalytic function
then so is $1/f$.
\end{proof}


\end{document}